\documentclass{jfm}
\usepackage[dvips]{graphicx}    \usepackage{psfrag}
\begin{document} 
\title[Stability analysis of rapid granular chute 
flows: formation of longitudinal vortices.]
{Stability analysis of rapid granular chute flows: 
formation of longitudinal vortices.}

\author[Y. Forterre and O. Pouliquen] {Y\ls O\ls \"E\ls L\ns \ns F\ls 
O\ls R\ls T\ls E\ls R\ls R\ls E\ns \and
 O\ls L\ls I\ls V\ls I\ls E\ls R\ns \ns P\ls O\ls
U\ls L\ls I\ls Q\ls U\ls E\ls N}

\affiliation{IUSTI, Technop\^ole Ch\^ateau-Gombert, 5 rue Enrico 
Fermi, 13453 Marseille Cedex 13, France.} 
\date{2001} 
\maketitle

\begin{abstract}
In a recent article (\cite{forterre01}), we have reported a new 
instability observed in rapid granular flows down inclined planes that 
leads to the spontaneous formation of longitudinal vortices.  From the 
experimental observations, we have proposed an instability mechanism 
based on the coupling between the flow and the granular temperature in 
rapid granular flows.  In order to investigate the relevance of the 
proposed mechanism, we perform in the present paper a 
three-dimensional linear stability analysis of steady uniform flows 
down inclined planes using the kinetic theory of granular flows.  We 
show that in a wide range of parameters, steady uniform flows are 
unstable under transverse perturbations.  The structure of the 
unstable modes are in qualitative agreement with the experimental 
observations.  This theoretical analysis shows that the kinetic theory 
is able to capture the formation of longitudinal vortices and 
validates the instability mechanism.
\end{abstract}

\section{Introduction}

Unlike classical fluids which are well-described by Navier-Stokes
constitutive equations, granular flows still lack an unifying 
description. For slow deformations at high density, multi-body 
interactions and friction between grains control the dynamics of the 
flow.  On the other hand, when the energy 
injected into the material is large, the particles are strongly 
agitated and interact mainly {\it via} instantaneous collisions. In 
this collisional regime, the material can be compared to a gas and a 
``granular temperature'' can be defined in relation with the random 
motion of the particles (\cite{ogawa80}; \cite{campbell90}). This 
analogy between a collisional granular flow and a molecular gas has 
led to the development of a kinetic theory for rapid granular flows 
(Jenkins \& Savage 1983; \cite{haff83}; \cite{lun84}; 
\cite{brey98}; \cite{sela98}) which is inspired by the kinetic 
theory of dense gas (\cite{chapman70}). However, the main difference 
with classical molecular gases is that the collisions between 
granular particles are inelastic. If no energy is supplied to the 
system, the granular temperature decays rapidly because each 
collision removes kinetic energy from the particles. It can be shown 
that a free dissipative gas can form dense clusters 
and eventually collapses in a finite time, as a consequence of 
inelasticity (\cite{mcnamara92}; \cite{goldhirsch93}). In order to 
maintain the collisional regime, energy must therefore be supplied to 
the system, for instance by strongly shaking the boundaries 
(Warr {\it et al} 1995; Falcon {\it et al} 1999; \cite{rouyer00}).
 In a rapid granular flow, another way to maintain 
granular temperature is to impose a shear to the mean flow. In 
that case, the granular temperature results from a balance between 
the production by the shear work and the lost due to the inelastic 
collisions. In return, the granular temperature influences the mean 
flow because the pressure and the transport coefficients (e.g. 
viscosity, thermal conductivity) depend on the
temperature, as in a molecular gas.  The coupling between the 
granular temperature and the mean flow is one of the fundamental 
properties of rapid granular flows. Understanding the role of this 
coupling is thus important in order to better define the analogies 
and the differences between classical fluid  and granular flows.

\begin{figure}
  \centering   \psfrag{z}{\hspace{0mm} $z$}
  \psfrag{x}{\hspace{-1mm} $x$}
  \psfrag{y}{\vspace{0mm} $y$}
  \psfrag{th}{\vspace{0mm} $\theta$}
  \psfrag{la}{\vspace{0mm} {\large$ \lambda$}}
  \psfrag{ho}{\vspace{0mm} $h_{g}$}
  \psfrag{h}{\vspace{0mm} {\large  $h$}}
  \psfrag{a}{\vspace{0mm} $(a)$}
  \psfrag{b}{\vspace{0mm} $(b)$}
  \includegraphics[scale=0.7]{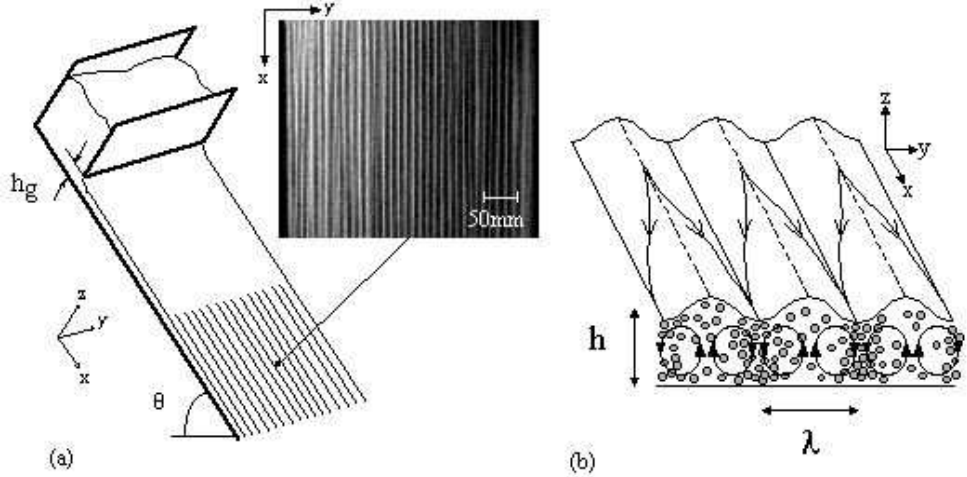}
  \caption{(a) Experimental set-up. The picture is a top view of the 
free surface lit from side when the instability is fully developed   
(the granular material is sand $0.25$ mm in mean diameter, the angle 
of inclination $\theta$ of the plane is $\theta = 41^{\circ}$, the opening 
 $h_{g}$ of the reservoir is  $h_{g}=13$ 
mm).   (b) Sketch of the flow deduced from the measurements showing 
the surface deformation, the longitudinal vortices and the density 
variations.}
  \label{fig:exp}
  \end{figure}

Recently, we have reported an instability observed in rapid granular 
flows down rough inclined planes which seems to result from this 
coupling (\cite{forterre01}). The experimental set-up was a rough 
inclined plane as sketched in figure \ref{fig:exp}(a). For high 
inclinations and large openings of the reservoir, the free surface 
deforms in a very regular pattern of longitudinal streaks parallel to 
the flow direction (see the picture in figure \ref{fig:exp}a). Velocity measurements 
of the grains have revealed that the streaks correspond to the 
formation of longitudinal vortices as sketched in figure 
\ref{fig:exp}(b).
Although such structures are common in fluid mechanics (e.g. Gšrtler 
vortices, see \cite{saric94} or  streaks in turbulent boundary layers, 
see \cite{kachanov94}), they had not been observed before in granular flows. 
The main experimental observations are the following:
\begin{itemize}
	\item  The wavelength $\lambda$ of the surface deformation scales 
	with the mean 
	thickness $h$ of the flow ($\lambda \sim 2-3\,h$).

	\item  The troughs correspond to downward part of the flow while the 
	crests correspond to upward part of the flow (see figure \ref{fig:exp}b). 

	\item  The $x$-component of velocity is greater in the troughs than in the 
	crests.

	\item  Troughs are dense whereas crest are dilute (see figure \ref{fig:exp}b). 

	\item  The pattern drifts slowly in the transverse direction.
\end{itemize}
These  observations suggest  the 
following instability mechanism to explain the formation of the 
longitudinal vortices (\cite{forterre01}). Because of the collisions with the rough 
bottom, particles close to the plane are strongly agitated i.e. the 
granular temperature is high at the bottom. Since high temperature 
means low density, density may become lower at the bottom than at 
the free surface. The flow is then mechanically unstable under 
gravity yielding longitudinal vortices. The mechanism we propose is 
analogous to the classical Rayleigh-B\'{e}nard instability observed when 
a fluid is heated from below. However, in the present case the 
heating is not imposed by a thermostat but is created by the flow 
itself through the coupling between temperature and shear specific 
to granular media. 

The above explanation for the vortices formation is based on  
density profile inversion.  In a granular dissipative gas, the 
density profile actually results from a complex balance between 
gravity, collisions and dissipation and its prediction is not 
straightforward.  Density profiles with higher density at the free 
surface than below have been
observed in numerical simulations of rapid granular flows using 
discrete element methods (\cite{campbell85}; \cite{azanza98}).  
However, no instability was observed because the simulations were 
two-dimensional.
 In the present paper, we want to investigate the relevance of the 
proposed instability mechanism by using the kinetic theory of 
granular flows. We present a  linear stability 
analysis of steady uniform flows down inclined planes, in the 
framework of the classical kinetic theory of Lun {\it et al} (1984). 
This theory provides a set of hydrodynamic equations coupling the 
density, the velocity and the granular temperature under the 
assumption of instantaneous binary inelastic collisions. Although 
steady uniform flows down inclined planes have been studied within 
this framework (\cite{anderson92}; \cite{ahn92}; \cite{azanza99}), 
no stability analysis has been performed. Stability studies of 
rapid granular flows with the kinetic theory have mainly focused 
on two-dimensional Couette flows (\cite{savage92}; \cite{wang97};
 Alam \& Nott 1998; \cite{nott99}).

We do not expect from the present study a complete and quantitative 
description of the longitudinal vortices instability,
since the applicability of the kinetic theory is known to be 
limited.  The lack of separation between the microscopic and 
macroscopic scales inherent to inelastic gases (\cite{tan99}) and the 
existence of multibody interactions when density is high, are serious 
difficulties which are the subject of active researches 
(\cite{goldhirsch99}).  However, the kinetic theory qualitatively 
contains the most important character of rapid granular flow i.e. the 
coupling between shear flow and particle agitation due to the 
inelasticity of collisions. This coupling being the core of the 
instability mechanism we propose, this analysis should
 capture at least qualitatively the formation of longitudinal 
vortices.

The paper is structured as follow.
 The Section 2 gives the governing equations and the boundary 
conditions we will use in the paper. Basic flows i.e. 
steady uniform flows down inclined planes are studied 
in Section 3. Section 4 is devoted to the linearization of the 
equations around the basic state and the numerical method. The 
results of the stability analysis are 
given in Section 5. Comparison with the experiment 
and discussion are presented in Section 6. Concluding remarks are 
given in Section 7.

\section{Kinetic theory of granular flows}
In this section, we recall the equations of the kinetic theory of 
granular flow we will use to investigate the formation of 
longitudinal vortices.  
\subsection{Governing  equations}

The kinetic theory of granular flows provides a set of 
hydrodynamic equations coupling the density $\rho$, the mean velocity 
$\bf u$, and the granular temperature $T$ under the assumption of 
instantaneous binary inelastic collisions (the granular temperature 
$T$ is defined by $\frac{1}{3}<{\delta \mathbf{u}}^2>$, where  
$\mathbf{\delta u}$ is the random velocity fluctuations). In the presence of 
gravity, the hydrodynamic equations are:  
\begin{eqnarray}
	\frac{\mathrm{d}\rho}{\mathrm{d}t} & = & -\rho \mathbf{\nabla .u},
	\label{eq:masscons}  \\
	\rho \frac{\mathrm{d}\mathbf{u}}{\mathrm{d}t} & = &  \rho 
\mathbf{g}+ \mathbf{\nabla .\Sigma},
	\label{eq:pcons}  \\
	\frac{3}{2} \rho \frac{\mathrm{d}T}{\mathrm{d}t} & = & 	
\mathbf{\Sigma \!:\!\nabla u} - \mathbf{\nabla .q}-\gamma,
	\label{eq:energycons}
\end{eqnarray}
where $\mathrm{d}/\mathrm{d}t = \partial / \partial t + 
\mathbf{u}.\mathbf{\nabla}$. The first equation (\ref{eq:masscons}) 
is the conservation of mass. The second equation (\ref{eq:pcons}) 
comes from the  conservation of momentum, where   $\mathbf{\Sigma}$ 
is the stress tensor and $\mathbf{g}$ is the gravity acceleration. 
The third equation (\ref{eq:energycons}) is the
energy equation where the temporal variation of the random kinetic 
energy is balanced by three terms. The term 
$\mathbf{\Sigma :\nabla u}$
represents the production of fluctuation energy due to the work of 
the stress $\mathbf{\Sigma}$ during shear. The term  $-\mathbf{\nabla .q}$, 
where  $\mathbf{q}$ is the flux of fluctuation energy, represents
 the conduction term. The term $\gamma$ is the collisional rate of energy 
dissipation.  The distinctive feature of rapid granular flows lies in 
the dissipative term $\gamma$ due to the inelastic collisions 
between particles. This last term is responsible for the coupling 
between the mean flow and the granular temperature.\\
 The kinetic theory gives the constitutive
relations for  $\mathbf{\Sigma}$, $\mathbf{q}$ and $\gamma$ as a 
function of the flow variables $\rho$, $\bf u$ and $T$. For the 
present purpose, we will use the closure due to Lun {\it et al} 
(1984). The total stress tensor $\mathbf{\Sigma}$, the heat flux 
 $\mathbf{q}$ and the  rate of energy 
dissipation $\gamma$ are written as:
         \begin{eqnarray}
         	\mathbf{\Sigma} & = &-\left\{  P(\nu,T)
         	- \xi(\nu,T)\mathbf{\nabla          	.u}\right\}\mathbf{I} 
+ 2\eta(\nu,T)\mathbf{S},
    \label{eq:stress}\\
    \mathbf{q} & = & -K(\nu,T)\mathbf{\nabla}T,
    \label{eq:fluxchaleur}\\
    \gamma & = &    
\frac{{\rho}_{p}}{d}(1-e^2)f_{5}(\nu)T^{\frac{3}{2}}.
   \label{eq:gamma}
         \end{eqnarray}
Here $\mathbf{I}$ is the identity matrix, $\mathbf{S} = 
\frac{1}{2}(u_{ij}+u_{ji}) - \frac{1}{3}u_{kk}{\delta}_{ij}$ is the 
deviatoric part of the rate of deformation, ${\rho}_{p}$ is the particle density, $\nu = \rho /{\rho}_{p}$ 
is the solid fraction and $d$ 
is the
 particle diameter.
We have omitted in (\ref{eq:fluxchaleur}) the contribution of the 
gradient of  compacity to the heat flux.  This term 
increases considerably the algebra while it has little effect on the 
qualitative nature of the results. Note that the collisional rate of 
energy dissipation $\gamma$ is proportional to $(1-e^2)$, where $e$ 
is the coefficient of inelasticity of the particles ($0<e\leq 1$). As 
in classical dense gases, the pressure $P(\nu,T)$, the viscosities 
($\eta(\nu,T)$, $\xi(\nu,T)$) and  the thermal conductivity
$K(\nu,T)$ depend on the solid fraction $\nu$ and the  temperature 
$T$:
\begin{equation}
 \left\{ \begin{array}{lll}
               P(\nu,T) & = & {\displaystyle {\rho}_{p}f_{1}(\nu)T},  
\\
    \eta(\nu,T) & = & {\displaystyle{\rho}_{p}d\, 
f_{2}(\nu)T^{\frac{1}{2}}},\\
    \xi(\nu,T)  & = & {\displaystyle{\rho}_{p}d\,     
f_{0}(\nu)T^{\frac{1}{2}}},\\
    K(\nu,T) & = & {\displaystyle{\rho}_{p}d\, 
f_{3}(\nu)T^{\frac{1}{2}}},
    \end{array}
    \right.
    \label{def:transport}
 	\end{equation}
where the dimensionless functions $f_{0}(\nu)$, $f_{1}(\nu)$, 
$f_{2}(\nu)$, $f_{3}(\nu)$ and $f_{5}(\nu)$ are given in table 
\ref{constitutivesfunctions} (\cite{lun84}).
\begin{table}
	\centering
	\begin{eqnarray*}
	f_{0}(\nu) & = & \frac{8\sqrt{\pi}}{3}\eta {\nu}^2 g_{0}(\nu),\\
	f_{1}(\nu)  & = & \nu + 4\eta{\nu}^2 g_{0}(\nu),\\
	f_{2}(\nu)  & = & \frac{5\sqrt{\pi}}{96}\left[ \frac{1}{\eta 	
(2-\eta)}\frac{1}{g_{0}(\nu)} + \frac{8}{5}\frac{3\eta -1}{2-\eta}\nu 
+ 	\frac{64}{25}\eta \left(\frac{3\eta -2}{2-\eta} + 	
\frac{12}{\pi}\right){\nu}^2  	g_{0}(\nu) \right],\\
	f_{3}(\nu)  & = & \frac{25\sqrt{\pi}}{16 \eta (41 - 
33\eta)}\left[\frac{1}{g_{0}(\nu)} +
	 \frac{12}{5}\eta (1+\eta (4\eta - 3))\nu +  	\frac{16}{25}{\eta}^2 
\left(9\eta (4\eta -3)  + 	\frac{4}{\pi}(41 - 33\eta)\right){\nu}^2  	
g_{0}(\nu) \right],\\
%	f_{4}(\nu)  & = & \frac{15\sqrt{\pi}}{4} \frac{(2\eta -1)(\eta - %	
%1)}{41 - 33\eta}\left( \frac{1}{\nu g_{0}(\nu)} + \frac{12}{5}\eta %	
%\right) \frac{\mathrm{d}}{\mathrm{d}\nu}({\nu}^2 g_{0}(\nu)),\\
	f_{5}(\nu)  & = & \frac{12}{\sqrt{\pi}} {\nu}^2 	g_{0}(\nu),\\
	f_{6}(\nu)  & = &  \frac{\pi \sqrt{3}}{6{\nu}_{m}} \nu 	g_{0}(\nu) 
,\\
	f_{7}(\nu)  & = &  \frac{ 3\sqrt{3}\pi}{12{\nu}_{m}} \nu 	g_{0}(\nu) 
,
	\label{}
\end{eqnarray*}
	\caption{Dimensionless constitutive functions. $\eta = 	
\frac{1}{2}(1+e)$}
	\label{constitutivesfunctions}
\end{table}
These functions involve the radial distribution function 
$g_{0}(\nu)$.  In the following, we shall used the one suggested by 
Lun and Savage (1986):
\begin{equation}
         	g_{0}(\nu) =          
\frac{1}{\left(1-\frac{\nu}{{\nu}_{m}}\right)^
         { 2.5{\nu}_{m}}},
         	\label{radialdistribution}
         	\end{equation}
where $\nu_{m}$ is the maximum solid fraction (${\nu}_{m} = 0.6$ in 
the following). This radial distribution function is suitable for 
free surface flows since the resulting equations have no singularity 
at $\nu = 0$. We have checked that this choice does not change 
qualitatively the results by using other kind of radial distribution 
functions, 
such as the Carnahan-Stirling radial distribution (\cite{jenkins83}). 
Finally, all the equations that follow will
 be written in term of  non-dimensional variables defined by:
\begin{equation}
         	(\tilde{x},\tilde{y}, \tilde{z})  =            	
\frac{1}{d}(x,y,z),\;\tilde{t}  =  \sqrt{\frac{d}{g}}\,t,\;
         	\tilde{\nu} = \frac{\rho}{\rho_{p}},\;\tilde{\mathbf{u}}  
=  \frac{1}{\sqrt{gd}}\,\mathbf{u},\;
         	\tilde{T}  = \frac{1}{gd}\,T.
         	\label{def:adim}
         \end{equation}
For  sake of simplicity in the notation, the tildes will be omitted 
and the solid fraction $\nu$ will be called density in the following 
.    
   \subsection{Boundary conditions}

In order to solve the problem of granular flows down rough inclined 
planes, we have to specify boundary conditions for $\nu$, 
$\mathbf{u}$ and $T$ both at the free surface of the flow and at the 
plane. Unlike classical fluid, the velocity in rapid granular flows 
does not vanish at a fixed solid boundary. Therefore, the rough plane 
may act as a source (resp. a sink) of fluctuating energy whether the 
shear work of the slip velocity is larger (resp. smaller) than the 
inelastic loss due to collisions with the plane.  Boundary conditions 
for rapid granular flows at a rough surface have been the subject of 
extensive researches (Hui {\it et al} 1984;  Jenkins \& Richman 
1986; Johnson {\it et al} 1990; Goldhirsch 1999).  Here we will 
use the heuristic approach of Johnson {\it et al} (1990)
relating the tangential stresses 
$\mathbf{t}.\mathbf{\Sigma}.\mathbf{n}$ and the heat flux 
 $\mathbf{q}.\mathbf{n}$ at the plane to density,  slip 
velocity $\mathbf{{u}_{s}}$ and  temperature by:
\begin{eqnarray}
			\mathbf{t}.\mathbf{\Sigma}.\mathbf{n} & = & 			
\eta^{\ast}(\nu,T)\left\|\mathbf{{u}_{s}}\right\|,\hspace{5mm}\mbox{at 			
the plane}, \label{bc:stress} \\
			\mathbf{q}.\mathbf{n} & = & 			
(\mathbf{u}.\mathbf{\Sigma}).\mathbf{n} - \gamma 			
^{\ast}(\nu,T),\hspace{5mm}\mbox{at the plane}.
			\label{bc:energy}
			\end{eqnarray}
The vector $\mathbf{t}$ is  parallel to the plane and defined as 
$\mathbf{t} = \mathbf{{u}_{s}}/\left\|\mathbf{{u}_{s}}\right\|$. The 
vector  
$\mathbf{n}$ is  normal to the plane. Finally the function 
$\eta^{\ast}(\nu,T)$ and $\gamma ^{\ast}(\nu,T)$ are given by:
\begin{equation}
	\left\{ \begin{array}{lll} \eta^{\ast}(\nu,T) & = & \phi 	f_{6}(\nu) 
T^{\frac{1}{2}}\\[1mm]

   \gamma ^{\ast}(\nu,T) & = & (1-{e_{w}}^2) f_{7}(\nu)    
T^{\frac{3}{2}},
	\end{array}
	\right.
	\label{def:transportbc}
	\end{equation}
where the dimensionless functions $f_{6}(\nu)$ and $f_{7}(\nu)$ are 
given in table \ref{constitutivesfunctions}. The relation 
(\ref{bc:stress}) is a  transfer of momentum balance at the plane. 
Equation (\ref{bc:energy}) expresses that heat can be produced at the 
plane if the shear work is stronger than the loss of energy due to 
 collisions with the plane.  The boundary conditions 
(\ref{bc:stress}) and (\ref{bc:energy}) depend on two dimensionless 
parameters, $\phi$ and $e_{w}$, which are related to the wall 
properties. The parameter $e_{w}$ is the particle-wall coefficient of 
restitution.  In the following, $e_{w}$ will be taken equal to the 
particle-particle coefficient of restitution $e$.
 The parameter $\phi$ is related to the rate of momentum transfer to 
the flow by the collision with the plane.  Its value can be connected 
to the wall geometry in the case of 2D-flows of disk 
(\cite{jenkins86}).  For a rough plane made of close-packed disks,
 one obtains a value $\phi \sim 0.1$.
    For 3D-flows, one can 
expect lower values of $\phi$ since collisions do not always occur
 in the shear plane. We will use in the following the value 
$\phi = 0.05$ and will discuss its influence later.

 At the free 
surface, the stresses and the energy flux must vanish.  However, the 
location of the free surface is not known {\it a priori} and its 
definition is somewhat artificial for very dilute flows.  Rather than 
define the location of the free surface, we impose stresses and 
energy flux to vanish when the distance from the plane goes to 
infinity. 
 It should be notice that the boundary conditions used here are 
somewhat different than those used in previous studies 
(\cite{johnson90}; \cite{anderson92};  \cite{ahn92}; 
\cite{azanza99}).   In the work of Anderson \& Jackson (1992), the 
thickness $h$ of the granular layer is taken as a control parameter. 
Numerical difficulties then arise when trying to match the 
stress-free condition at the free surface (\cite{johnson90}).  Ahn 
{\it et al} (1992) do not define the free surface but impose 
stresses to vanish at infinity.   However, they arbitrarily fixe the 
density, velocity and 
granular temperature at the plane.  The same procedure is used by 
Azanza {\it et al} (1999).  Here, we adopt a mixed point of view 
since at the plane the boundary conditions of Johnson {\it et al} 
(1990) are satisfied whereas at infinity the procedure of  Ahn {\it 
et al} (1992) is chosen.\\\\

\section{Steady uniform flows}

The first step in order to perform a linear stability analysis is to 
determine the basic flow i.e. two-dimensional steady uniform flows.  
Steady uniform flows down inclined planes have already been studied 
in the framework of the kinetic theory (Anderson \& Jackson 1992; 
\cite{ahn92}; \cite{azanza99}).  It is not in the scope of the 
present study to make an extensive investigation of steady uniform 
flows.  Rather, we shall focus on the shape of the density profile, 
which plays an important
role in  the instability mechanism we propose.
\subsection{Equations for steady uniform flows}
We apply equations (\ref{eq:masscons})-(\ref{eq:energycons})
 with boundary conditions (\ref{bc:stress}) and  
(\ref{bc:energy}) to two-dimensional steady uniform flows down  
inclined planes. We thus look for solution for density, velocity 
and temperature in the following form:
 \begin{equation}
\left\{ \begin{array}{ccc}
	\nu (x,y,z,t) & = & {\nu}_{0}(z),\\
	\mathbf{u}(x,y,z,t) & = & {u}_{0}(z) \mathbf{x},\\
	 T(x,y,z,t) & = & T_{0}(z).
	\end{array}
	\right.
	\label{def:T0}
\end{equation}
In such a flow the derivatives parallel to the plane are zero and the 
mass-conservation equation (\ref{eq:masscons}) is 
satisfied. The momentum-conservation equation (\ref{eq:pcons}) in 
the flow direction ($x$-direction) and in the direction normal to the 
plane ($z$-direction) becomes:
 \begin{eqnarray}
           	\frac{\mathrm{d}{{\Sigma}_{xz}}_{0}}{\mathrm{d}z} & = 
&          	-{\nu}_{0}\sin \theta,          	\label{eq0:ox} \\
         	\frac{\mathrm{d}{{\Sigma}_{zz}}_{0}}{\mathrm{d}z} & = 
&          	{\nu}_{0}\cos \theta,
         	\label{eq0:oz}
         \end{eqnarray}
where $\theta$ is the angle of inclination of the plane, 
${{\Sigma}_{xz}}_{0} = f_{2}({\nu}_{0}) 
{T_{0}}^{1/2}\mathrm{d}{u}_{0}/\mathrm{d}z $ and ${{\Sigma}_{zz}}_{0} 
= - f_{1}({\nu}_{0}) T_{0}$. The energy-conservation 
(\ref{eq:energycons}) simplifies to:
\begin{equation}
         0 = {{\Sigma}_{xz}}_{0} 
\frac{\mathrm{d}{u}_{0}}{\mathrm{d}z} -          
\frac{\mathrm{d}{q_{z}}_{0}}{\mathrm{d}z} - \gamma_{0},
         	\label{eq0:energy}
         	\end{equation}
where ${q_{z}}_{0} = - f_{3}({\nu}_{0}) 
{T_{0}}^{1/2}\mathrm{d}T_{0}/\mathrm{d}z$ and $\gamma_{0} = (1-e^2) 
f_{5}({\nu}_{0}){T_{0}}^{3/2}$.\\
From (\ref {eq0:ox}) and  (\ref {eq0:oz}) together with boundary 
conditions ${{\Sigma}_{xz}}_{0}={{\Sigma}_{zz}}_{0}=0$ at infinity, 
one obtains ${{\Sigma}_{xz}}_{0} = -\tan \theta 
\;{{\Sigma}_{zz}}_{0}$. Therefore, equations (\ref{eq0:ox}), 
(\ref{eq0:oz}) and (\ref{eq0:energy}) can be written in the following 
form:
\begin{eqnarray}
			\frac{\mathrm{d}{\nu}_{0}}{\mathrm{d}z} & = & 			
-\frac{1}{{f_{1}}^{'}({\nu}_{0})T_{0}} \left( {\nu}_{0}\cos \theta 
+ 			f_{1}({\nu}_{0})\frac{\mathrm{d}T_{0}}{\mathrm{d}z}\right), 			
\label{eq0:nu0}\\ 			\frac{\mathrm{d}u_{0}}{\mathrm{d}z} & = & 			
\frac{f_{1}({\nu}_{0})}{f_{2}({\nu}_{0})} \tan \theta 			
{T_{0}}^{\frac{1}{2}},
			\label{eq0:u0}\\ 			\frac{{\mathrm{d}}^2 T_{0}}{\mathrm{d}z^2} & = 
& 			\left(\frac{(1-e^2)f_{5}({\nu}_{0}) f_{2}({\nu}_{0}) 			
-{f_{1}}^2({\nu}_{0}) \tan \theta 			
}{f_{2}({\nu}_{0})f_{3}({\nu}_{0})}\right)T_{0}+
			 \frac{{\nu}_{0}{f_{3}}'({\nu}_{0})} 			
{f_{3}({\nu}_{0}){f_{1}}'({\nu}_{0}) T_{0}}\cos 			\theta 
\frac{\mathrm{d}T_{0}}{\mathrm{d}z}\nonumber \\
			& &           	
\hspace{50mm}+\frac{1}{T_{0}}\left(\frac{f_{1}({\nu}_{0}){f_{3}}'({\nu}_{0})}
         	{f_{3}({\nu}_{0}){f_{1}}'({\nu}_{0})}
			-\frac{1}{2}\right){\left(\frac{\mathrm{d}T_{0}}{\mathrm{d}z}\right)}^2,
		    \label{eq0:T0}
\end{eqnarray}
where the dashes in ${f_{1}}'$ and ${f_{3}}'$ denote differentiation 
with respect to ${\nu}_{0}$.\\
This system of three nonlinear ordinary differential equations 
requires boundary conditions both at the plane and at infinity.  At 
infinity, the stresses and the flux of energy must vanish.  It can be 
shown (\cite{ahn92}) that these conditions are satisfied if and only if the 
temperature gradient 
 vanishes at infinity: 
\begin{equation}
         \frac{\mathrm{d}T_{0}}{\mathrm{d}z}\rightarrow 0          
\hspace{5mm}\mbox{when  $z\rightarrow \infty$}.
         	\label{bc0:infini}
\end{equation}
At the plane, the relations (\ref {bc:stress}) and (\ref{bc:energy}) 
become:
\begin{eqnarray}
f_{2}({\nu}_{0}){{T}_{0}}^{\frac{1}{2}}\frac{\mathrm{d}{u}_{0}}{\mathrm{d}z} 			
& = & \phi f_{6}({\nu}_{0}) 			
{{T}_{0}}^{\frac{1}{2}}{u}_{0}\hspace{5mm}\mbox{at $z=0$}, 
\label{bc0:stress} \\
-f_{3}({\nu}_{0}) 			
{{T}_{0}}^{\frac{1}{2}}\frac{\mathrm{d}T_{0}}{\mathrm{d}z} 			& = & 
f_{2}({\nu}_{0}) {{T}_{0}}^{\frac{1}{2}} 			
\frac{\mathrm{d}{u}_{0}}{\mathrm{d}z}{u}_{0} - 			
(1-e^2)f_{7}({\nu}_{0}) 			
{{T}_{0}}^{\frac{3}{2}}\hspace{5mm}\mbox{at $z=0$.}
			\label{bc0:energy}
\end{eqnarray}
These relations can be combined with equation (\ref{eq0:u0}) to 
express the temperature and the gradient of temperature at the plane 
as a function of density and velocity at the plane:
\begin{eqnarray}
			T_{0} & = & \frac{{\phi}^2 			
{f_{6}}^2({\nu}_{0})}{{f_{1}}^2({\nu}_{0}) {\tan}^2\theta 			} 
{{u}_{0}}^2\hspace{5mm}\mbox{at $z=0$}, \label{bc0:temperature} \\
			\frac{\mathrm{d}T_{0}}{\mathrm{d}z} & = & \frac{\phi 			
f_{6}({\nu}_{0})}{f_{3}({\nu}_{0})}\left( -1 + \frac{\phi 			
f_{6}({\nu}_{0}) (1-e^2)f_{7}({\nu}_{0})}{ 			{f_{1}}^2({\nu}_{0}) 
{\tan}^2 \theta}\right) {{u}_{0}}^2 			\hspace{5mm}\mbox{at $z=0$.}
			\label{bc0:gradienttemperature}
\end{eqnarray}
 In order to integrate numerically the above boundary value problem, 
a  ``shooting method'' has been chosen.  The angle $\theta$ and the 
density at the wall ${\nu}_{0}(0)$ are the two control parameters 
whereas the slip velocity $u_{s}={u}_{0}(0)$ is taken as the shooting 
parameter.  Then, the temperature and the gradient of temperature at 
the plane are given by the equations (\ref{bc0:temperature}) and  
(\ref {bc0:gradienttemperature}).  The differential equations (\ref  
{eq0:nu0})-(\ref {eq0:T0}) can therefore be  
integrated from $z = 0$ to $z=\infty $.  This procedure is repeated 
for different values of the shooting parameter $u_{s}={u}_{0}(0)$
 until the boundary condition at 
infinity  $\mathrm{d}T_{0}/\mathrm{d}z = 0$ is satisfied. In practice,  a 
fourth-order Runge-Kutta method is employed and the equations are 
integrated up to a maximum value $z = z_{max}$. The boundary condition 
at infinity is considered to be satisfied when the gradient of 
temperature is  smaller than $10^{-5}$ at $z = z_{max}$.

\subsection{Results}

\begin{figure}
  \centering   \psfrag{h}{\hspace{0mm} $h$}
  \psfrag{nu}{\vspace{0mm} $\bar{\nu}$}
  \includegraphics[scale=0.6]{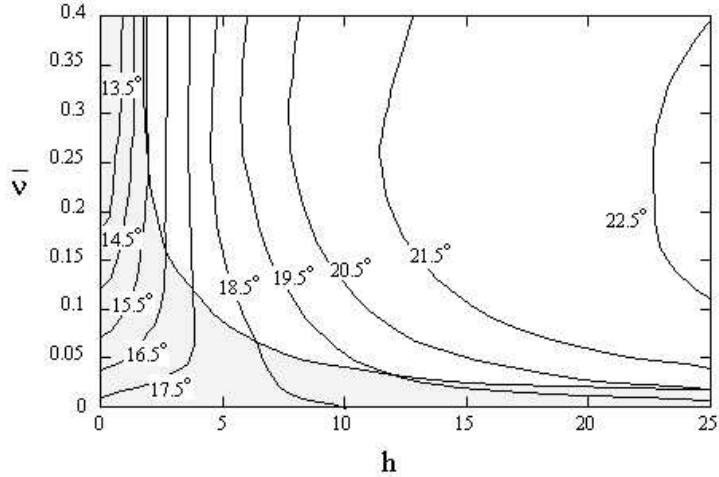}
  \caption{Phase diagram for steady uniform flows in the ($h$, 
$\bar{\nu}$)
  plane. The thin solid lines give the contours of constant angle 
$\theta$. The grey zone is the domain of non-inverted density 
profiles. ($e=0.6$, $\phi = 0.05$).}
  \label{fig:esudiag}
  \end{figure}

    By applying the above procedure to different values of the angle 
$\theta$
 and the density ${\nu}_{0}(0)$, we were able to explore the whole 
set  of steady uniform flows. However, in order to help the physical 
interpretation of the results, we  present the results in the ($h$, $\bar{\nu}$) plane
 where $h$ is the thickness of
 the flow and  $\bar{\nu}$ the mean density.   The thickness $h$ is 
defined {\it a posteriori} by the value $z=h$ for which the density 
is $1\%$
of the maximum density inside the flow.  The mean density $\bar{\nu}$ 
is simply given by $\bar{\nu} = (1/h)\int_{0}^{h} {\nu}_{0}(z) 
dz$. The relation between ($\theta$, ${\nu}_{0}(0)$)
and ($h$, $\bar{\nu}$) is univocal i.e. a single solution exists for a 
given set ($h$, $\bar{\nu}$) (which is not the case if one chooses the 
flow rate as a control parameter, see \cite{johnson90} and 
\cite{anderson92}). The relation between ($\theta$, ${\nu}_{0}(0)$) 
and  ($h$, $\bar{\nu}$) is shown in figure \ref{fig:esudiag} for 
$e=0.6$ and $\phi = 0.05$. In this figure,  the thin solid lines are 
obtained for constant inclination $\theta$ by varying 
${\nu}_{0}(0)$.     We observe that the angle $\theta$ has a strong 
influence on the thickness of the flow $h$, the thickness of the flow 
rapidly increases when increasing  the angle. For angles $\theta > 23^{\circ }$, 
the thickness diverges  whereas  for $\theta < 14^{\circ }$, the 
thickness becomes less than one grain diameter. 
Therefore, for  a given set of parameters ($e$, $\phi$) there exists a
finite  range of 
angles where steady uniform flows are obtained.  These results are 
consistent with analytical solutions for steady uniform flows found in 
the high-density limit (\cite{anderson92}).

\begin{figure}
  \centering   \psfrag{z}{\hspace{-1mm} $z$}
  \psfrag{u}{\vspace{3mm}\hspace{-5mm} $u_{0}(z)$}
  \psfrag{nu}{\vspace{2mm}\hspace{-5mm} $\nu_{0}(z)$}
  \psfrag{T}{\vspace{2mm}\hspace{-5mm} $T_{0}(z)$}
  \psfrag{a}{(a)}
  \psfrag{b}{(b)}
  \psfrag{c}{(c)}
  \psfrag{h}{\hspace{2mm}$h$}
  \psfrag{nub}{\hspace{2mm}$\bar{\nu}$}
  %scale original 0.6
  \includegraphics[scale=0.5]{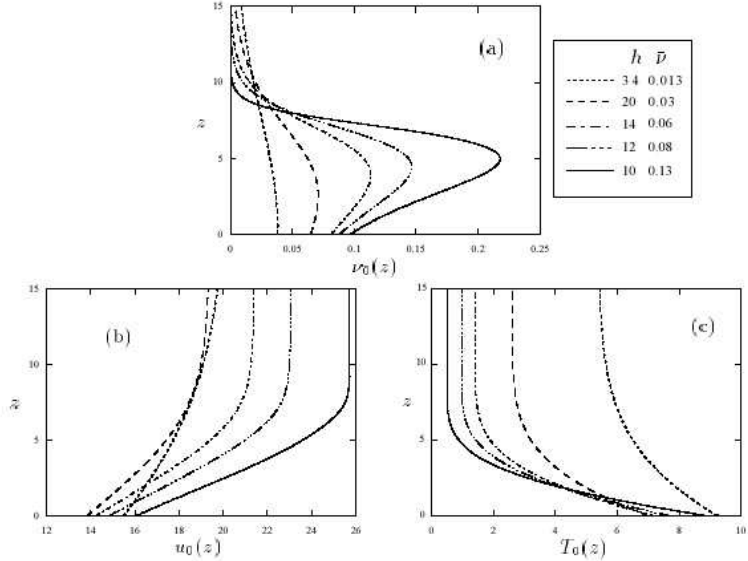}
  \caption{(a) Density profiles $\nu_{0}(z)$, (b) velocity profiles 
$u_{0}(z)$ and (c) temperature profiles $T_{0}(z)$ for steady uniform 
flows at a fixed angle   ($\theta=20.5^{\circ }$) when the mean 
density is increased. The inset gives the value of ($h$, $\bar{\nu}$) 
for the different   profiles ($e=0.6$, $\phi = 0.05$).}
  \label{fig:esu}
  \end{figure}

Typical solutions ($\nu_{0}(z), u_{0}(z), T_{0}(z)$) for steady 
uniform flows are given in figure \ref{fig:esu}. They are obtained for 
$\theta =20.5^{\circ }$ by increasing the mean density $\bar{\nu}$
i.e. along the thin solid line $\theta =20.5^{\circ }$ in figure \ref{fig:esu}.
 As the mean density is increased, the shear close to the 
plane increases (see figure \ref{fig:esu}b). The difference of temperature between 
the plane and the free surface also increases with the mean density
 (see figure \ref{fig:esu}c). 
Note that with our boundary conditions, the temperature is
always higher at the plane than at the free surface.
Finally, the density profile $\nu_{0}(z) $ is strongly modified when 
varying the mean density (see figure \ref{fig:esu}a).  For dilute 
flows, the thickness
 $h$ is large and the density profile is non-inverted i.e. the 
density  decreases with $z$.  As the mean density  
increases, the flow becomes thinner and the free surface is better 
defined.  At a given mean density, the density profile 
becomes inverted i.e. the maximum of density is no longer at the 
plane but in the bulk.   In the ($h$, $\bar{\nu}$) diagram 
of figure \ref{fig:esudiag}, the limit between non-inverted density 
profiles and inverted density profiles is given by the solid line. 
Above this line, the maximum of density is in the bulk.

The shape of density profiles can be understood qualitatively by 
inspecting the equations (\ref{eq0:nu0})-(\ref{eq0:T0}).  
Eq(\ref{eq0:nu0}) shows that the sign of the density gradient results 
from a competition between gravity, which tends to increase density 
close to the plane, and the negative temperature gradient, which 
tends to decrease density close to the plane.
 The temperature gradient results from 
the shearing at the base, which heats the material from below and the 
inelastic dissipation in the bulk, which cools the material.
Therefore, inverted density profiles are observed when heating due to
the shear 
at the base and the cooling due to dissipation in the bulk
create a temperature gradient strong enough to overcome the 
gravity.

\subsection{Influence of the parameters}

The results presented so far have been obtained using $e=0.6$ and 
$\phi = 0.05$.    Here we discuss briefly the 
qualitative influence of the inelastic coefficient $e$ and the 
boundary parameter $\phi$ on the solutions presented above.  First, 
inverted density profiles are observed only for small values of 
$\phi$.  For $\phi>0.3$, no inverted density profile can be obtained, 
neither by changing the control parameters ($h$, $\bar{\nu}$) nor by 
modifying the coefficient of inelasticity $e$. For high values of 
$\phi$, the 
plane is indeed no longer a source of fluctuating energy
 (see equation \ref{bc0:gradienttemperature}).  The density 
gradient close to the plane is therefore negative and no inverted 
density profile can be obtained. For values of $\phi$ less than 
$0.3$, the results are qualitatively the same as for $\phi = 0.05$.
Varying $\phi$ changes the range of inclinations where steady uniform
flows are observed.  For example, 
the maximum angle is increased from $\theta \sim 23^{\circ }$ to 
$\theta \sim 32^{\circ }$ when increasing$\phi$  from 0.05 to 0.15 
($e=0.6$). No qualitative change is also observed when changing the 
inelasticity $e$. Increasing the value of $e$ decreases the angles 
and shrinks the domain of steady uniform flows.  With $e=0.8$ 
($\phi=0.05$), the range of angles is $11^{\circ }<\theta <18^{\circ }$.

Steady uniform flows with inverted density profiles can therefore be 
obtained in the framework of the kinetic theory of granular flows.   
The question we address in this study is the stability of these 
inverted density profiles under transverse perturbations.  Obviously, 
gravity is a destabilizing effect since the heavy fluid is above the 
light one.  However, gravity has to overcome the stabilizing effects 
due to   viscosity and  thermal conductivity.  In a rapid granular flow, 
viscosity, thermal conductivity and density profiles are coupled 
to the flow and the prediction of the stability properties is not 
straightforward.

\section{Three-dimensional linear stability analysis}

%%%%%%%%%%%%%%%%%%%%%%%%%%%%%%%%%%%%%%%%%%%%%%%%%%%%%%%%%%%%%%%%%%%%%%%%%%%%%%%%%%%%%

%%%%%%%%%%%%%%%%%%%%%%%%%%%%%%%%%%%%%%%%%%%%%%%%%%%%%%%%%%%%%%%%%%%%%%%%%%%%%%%%%%%%%

We investigate the stability of steady uniform flows ($\nu_{0}$, 
$u_{0}$ $T_{0}$) found in the previous section using the classical 
normal mode analysis (\cite{drazin81}). The basic flow is perturbed 
by infinitesimal disturbances, and their time evolution are studied by 
linearizing the governing equations about the basic state. The 
perturbations are then decomposed into different Fourier modes and 
because of the linearity of the governing equations, the stability of each 
mode can be 
analyzed separately.

 \subsection{Normal mode analysis}
 As point out by Alam \& Nott (1998), the Squire theorem does not 
hold for rapid granular flows since density and temperature 
are not constant across the layer. Therefore, the first instability 
is not necessary two-dimensional. In this study, we are interested in the 
formation of longitudinal structures and we restrict our analysis to 
flows which are invariant along the $x$-direction. The  flow ($\nu$, 
$u$, $v$, $w$, $T$) is perturbed around the basic flow ($\nu_{0}$,
 $u_{0}$, $T_{0}$) and written as:
  \begin{equation}
   \begin{array}{lll}
    \nu  =  \nu_{0}(z)  +  \nu_{1}(y,z,t),& \left\{ \begin{array}{l}
    %\nu & = & \nu_{0}(z) & + & \nu_{1}(y,z,t)\\
    u  = u_{0}(z)  +  u_{1}(y,z,t)\\
    v  =   v_{1}(y,z,t)\\
    w  =    w_{1}(y,z,t)
    \end{array}
    \right.
    ,&T  =  T_{0}(z)  +  T_{1}(y,z,t),
    \end{array}
    \label{def:ordre1}
 \end{equation}
where ${\nu}_{1}$, $(u_{1}, 
v_{1}, w_{1})$ and $T_{1}$ are respectively the density, 
velocity and temperature  disturbances. By substituting (\ref{def:ordre1})
 into the  
governing equations (\ref{eq:masscons})-(\ref{eq:energycons}) and 
then  
linearizing about the basic state, we obtain a set of linear 
equations for $(\nu_{1}, u_{1}, v_{1}, 
w_{1}, T_{1})$ (see Appendix A.1). Then, we seek normal mode 
solutions for density, velocity and temperature 
perturbations:
 \begin{equation}
 	(\nu_{1}, u_{1}, v_{1}, w_{1}, T_{1}) = \Re \mathrm{e}\left[ \left( 
\hat{\nu}(z), \hat{u}(z),  \hat{v}(z), \hat{w}(z), 
\hat{T}(z)\right)\;\mathrm{e}^{ \sigma t + iky}\right].  	
\label{def:modes}
 \end{equation}
We have restricted the stability analysis to temporal stability i.e.
the growth rate $\sigma$ is complex whereas the  transverse 
wavenumber  $k$ is assumed to be real. The basic flow ($\nu_{0}$, 
$u_{0}$, $T_{0}$) is unstable under the transverse perturbation of 
wavenumber $k$ if the real part of the growth rate, $\Re e(\sigma)$, 
is positive. After some algebra, it can be shown that the perturbed 
variables satisfy a system of ordinary differential equations:
\begin{equation}
	L_{0}(z)\frac{\mathrm{d^2}}{\mathrm{d}z^2}\hat{X}(z) + 
M_{0}(z)\frac{\mathrm{d}}{\mathrm{d}z}\hat{X}(z)
	+ N_{0}(z)\hat{X}(z) = 0,
	\label{eq1:final}
	\end{equation}
where $\hat{X}(z)$ is the five-elements vector defined by $\hat{X}(z) 
= (\hat{\nu}(z), \hat{u}(z),  \hat{v}(z), \hat{w}(z), \hat{T}(z))$, 
and $L_{0}(z)$, $M_{0}(z)$,  $N_{0}(z)$ are  $5\times 5$ matrices 
which are given in Appendix A.2. Note that   these matrices depend 
on the basic flow ($\nu_{0}$, $u_{0}$, $T_{0}$), on the wavenumber 
$k$ and on the growth rate $\sigma$. The boundary conditions at the 
plane can be written in the same manner:
\begin{equation}
	Q_{0}(z)\frac{\mathrm{d}}{\mathrm{d}z}\hat{X}(z)
	+ R_{0}(z)\hat{X}(z) = 0\hspace{5mm} \mbox{at $z=0$,}
	\label{bc1plan:final}
	\end{equation}  where $Q_{0}(z)$ and $R_{0}(z)$ are matrices which 
are also given in Appendix A.2.
At infinity, the disturbances $\hat{X}(z)$ have to vanish.
 It can be shown by doing an asymptotic expansion of equations 
(\ref{eq1:final}) that the disturbances  ($\hat{u}$, 
$\hat{v}$, $\hat{w}$, $\hat{T}$)  decrease as $\mathrm{exp}(-kz)$ when $z$ is 
much larger than the characteristic thickness  of the basic flow (see 
Appendix B). Matching the boundary conditions at infinity thus leads 
to numerical difficulties when $k$ becomes small since the 
computational domain varies as $1/k$.  Instead of writing 
the boundary conditions at infinity, the equations are integrated up 
to a finite value $z=z_{max}$ and the disturbances ($\hat{u}$, 
$\hat{v}$, $\hat{w}$, $\hat{T}$) are assumed to satisfy  
\begin{equation}
  	\frac{\mathrm{d}}{\mathrm{d}z}\left( \hat{u}(z),  \hat{v}(z), 
\hat{w}(z), \hat{T}(z)\right) = -k \left( \hat{u}(z),  \hat{v}(z), 
\hat{w}(z), \hat{T}(z)\right),   	\hspace{5mm}\mbox{at $z=z_{max}$.}
  	\label{bc1infini:da}
  \end{equation}
The system of five ordinary differential equations (\ref{eq1:final}) 
together with the eight boundary conditions (\ref{bc1plan:final}) and 
(\ref{bc1infini:da})  constitute an eigenvalue problem. For a given 
basic flow ($\nu_{0}$, $u_{0}$, $T_{0}$) and wavenumber $k$, a 
non-zero solution ($\hat{\nu}$, $\hat{u}$, $\hat{v}$, $\hat{w}$, 
$\hat{T}$) exists only for specific values of the  growth rate 
$\sigma$.     
\subsection {Numerical method}

We have solved the above eigenvalue problem thanks to a Chebychev 
spectral collocation method (\cite{gottlieb84}). Chebychev 
collocation approach has shown to be well-adapted to the stability of 
boundary-layer flows, since Chebychev polynomials resolve the 
boundary regions extremely well (\cite{malik90}). It is then suitable 
for our flow  which is localized close to the plane.  Moreover, the 
use of collocation makes the derivatives easy to compute and 
simplifies the treatment of boundary conditions. \\
The principle of the Chebychev spectral collocation method is to 
discretize the ordinary differential equations (\ref{eq1:final}) by 
interpolating the perturbed functions ($\hat{\nu}(z)$, $\hat{u}(z)$, 
$\hat{v}(z)$, $\hat{w}(z)$, $\hat{T}(z)$) on $N+1$ collocation points 
given by
\begin{equation}
 {\varsigma}_{j}=\cos\frac{\pi j}{N};\hspace{10mm}(i = 0,\ldots,N).  	
\label{eq:xi}
\end{equation}
In order to relate the  Chebychev space ($\varsigma \in [-1,1]$) to the 
physical domain ($z \in [0,z_{max}]$), we use a two-parameters algebric 
transformation
(\cite{malik90})
\begin{equation}
 	z = a\frac{1+\varsigma}{b-\varsigma},  	\label{eq:dil}
 \end{equation}
where $b = 1+2a/z_{max}$ and $a = z_{h}z_{max}/(z_{max}-2z_{h})$. The 
location  
$z_{h}$  corresponds to $\varsigma = 0$, i.e. half of the 
grid points are located in the region $0\leq z \leq z_{h}$. This 
mapping allows to cluster grid points near the plane. Using the 
expression for the derivatives at the collocation points 
(\cite{gottlieb84}), the system of ordinary differential equations 
(\ref{eq1:final}) together with the  boundary conditions 
(\ref{bc1plan:final}) and (\ref{bc1infini:da}) reduce to
a linear algebraic eigenvalue problem that can be written in the 
following form: \begin{equation}
 	\mathcal {A}\hat{\mathcal{X}} = \sigma \mathcal{ 
B}\hat{\mathcal{X}}.  	\label{eq:valpr}
 \end{equation}
Here, $\hat{\mathcal{X}}$ is the vector containing the $5(N+1)$  
elements of the interpolation of ($\hat{\nu}(z)$, $\hat{u}(z)$, 
$\hat{v}(z)$, $\hat{w}(z)$, $\hat{T}(z)$) on  the ($N+1$) collocation 
points: \begin{equation}
\left({\hat{\nu}}_{0},\ldots,{\hat{\nu}}_{N};\;{\hat{u}}_{0},\ldots, 	
{\hat{u}}_{N};\;{\hat{v}}_{0},\ldots, 	
{\hat{v}}_{N};\;{\hat{w}}_{0},\ldots, 	
{\hat{w}}_{N};\;{\hat{T}}_{0},\ldots, {\hat{T}}_{N}\right)
	\label{eq:vector}
 \end{equation}
and $\mathcal{A}$ and $\mathcal{B}$ are $5(N+1)\times 5(N+1)$ matrices 
computed from the matrices $L_{0}$, $M_{0}$ and $L_{0}$.  
$\mathcal{A}$ and $\mathcal{B}$ depend on the basic flow and on the 
wavenumber $k$.  Since the boundary conditions (\ref {bc1infini:da}) 
do not contain the eigenvalue $\sigma$, the matrix $\mathcal{B}$ is 
singular. It must be noticed that the above discretization requires 
ten boundary conditions while the physical boundary conditions 
(\ref{bc1plan:final})
 and (\ref{bc1infini:da}) give
only eight relations for the perturbed fields (four at the plane and 
four at infinity). The same problem arises in the stability analysis 
of compressible hydrodynamic flows (\cite{malik90}).  Here we have 
chosen
the vertical momentum balance at the plane and the conservation of 
mass at infinity  as the  two extra boundary conditions. 

The generalized eigenvalue problem (\ref{eq:valpr}) is solved by 
using the scientific software MatLab.  The advantage of spectral 
methods is that the whole spectrum of eigenvalues may be obtained.  
However, many eigenvalues are spurious eigenvalues due to the 
discretization (\cite{mayer92}).  The location of these spurious modes 
are very sensitive to the number $N$ of collocation points while the 
physical modes are insensitive.  This allows us to select the few physical 
modes among the large spectrum of the discretized problem.  With a 
typical number of collocation points $N = 50$, the absolute error in 
the physical eigenvalues is less than $10^{-7}$. 
\section{Results}

\subsection{Diagram of stability}
\begin{figure}
  \centering   \psfrag{h}{\hspace{-5mm} $h$}
  \psfrag{nu}{\hspace{5mm} $\bar{\nu}$}
  \psfrag{stable}{stable}
  \psfrag{a}{A}
  \psfrag{b}{B}
  \psfrag{c}{C}   \psfrag{density profiles}{density profiles}
  \psfrag{stationary instability}{\vspace{0mm} stationary instability}
  \psfrag{oscillatory}{\vspace{0mm}   propagating}
  \psfrag{instability}{instability}
  %\psfrag{temperature}{\vspace{0mm} temperature}
  %scale original 0.7
  \includegraphics[scale=0.6]{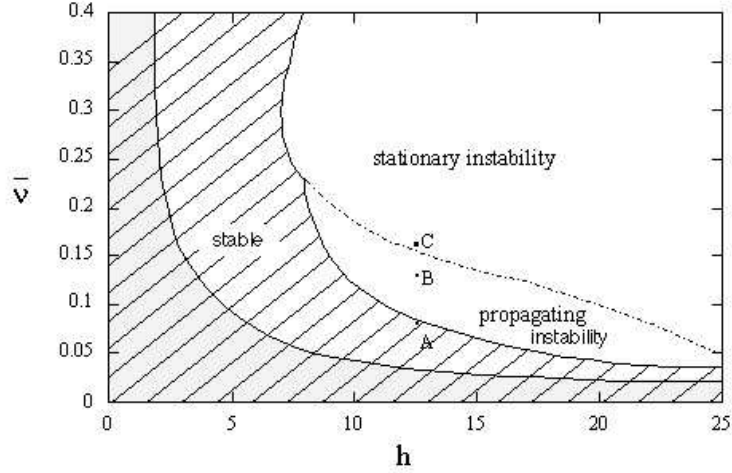}
  \caption{Stability diagram for $e=0.6$ and $\phi = 0.05$. The flow 
is stable inside the hatched region and unstable outside. The bold 
solid line is the marginal curve. The dotted line separates a region 
where the most amplified mode is propagating from a region where it is 
stationary. 
The grey zone is the domain of non-inverted density profiles.}
  \label{fig:diagstab}
  \end{figure}

 Figure \ref{fig:diagstab} gives the diagram of stability in the 
plane ($h$, $\bar{\nu}$) for $e = 0.6$ and  $\phi = 0.05$.  The 
region of non-inverted density profiles is presented on the same 
diagram. Since we are essentially interested in thin granular layer 
flows, we have limited our stability analysis to  $h$ < 30 particle 
diameters. The study is also limited to mean density less than  
$0.4$. For higher mean density, the density profile is very sharp 
and the number of collocation points 
required for accurate computation increases dramatically.

The stability diagram can be divided in three regions: a stable 
region (hatched region), an unstable region with stationary modes, an 
unstable region with propagating modes. The stable zone is delimited 
by the marginal curve where the real part of $\sigma$ vanishes.
One observes that for $h$ less than 7 particle diameters, the flow is 
always stable whatever the value of the mean density. The unstable 
region is composed of two regions: one where the most amplified mode 
is stationary i.e. $\Im$m($\sigma$)$=0$, and one where the most 
amplified mode is propagating i.e. with a non zero phase velocity 
($\Im$m($\sigma$)$\neq 0$). Propagating modes are obtained at low 
density while stationary instability takes place at high density.   

\begin{figure}
  \centering   \psfrag{k}{\vspace{3mm} $k$}
  \psfrag{kc}{$k_{c}$}
  \psfrag{sigmar}{\hspace{-6mm} \vspace{0mm} $\Re \mathrm{e}(\sigma)$}
  \psfrag{sigmai}{\hspace{-6mm} \vspace{0mm} $\Im \mathrm{m}(\sigma)$}
 \psfrag{a}{A}
 \psfrag{b}{B}
 \psfrag{c}{C}
 % \psfrag{v}{\hspace{0mm} $\hat{v}$}
 % \psfrag{w}{\hspace{0mm} $\hat{w}$}
 % \psfrag{t}{\hspace{0mm} $\hat{T}$}
 % \psfrag{nu}{\hspace{0mm} $\hat{\nu}$}
 %\psfrag{nu}{\vspace{0mm} $\bar{\nu}$}
 %scale original 0.9
  \includegraphics[scale=0.5]{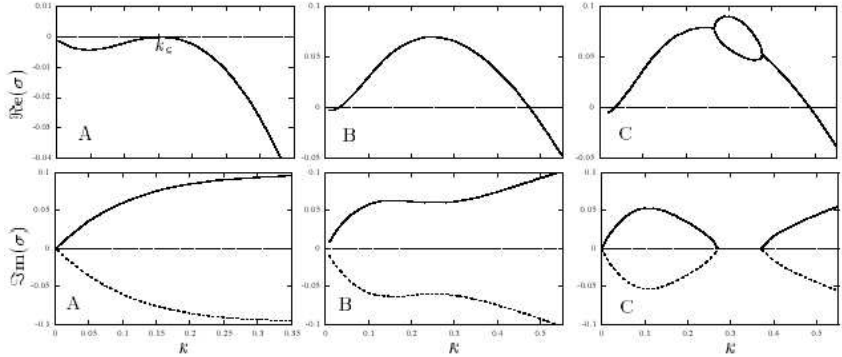}
  \caption{Growth rate $\Re \mathrm{e}(\sigma)$ and   
$\Im\mathrm{m}(\sigma)$ as a function of the wavenumber $k$   
for the most dangerous mode. The thickness is constant  ($h=12.8$) and 
the mean density is increased:  $\bar{\nu}=0.08$ (A),
   $\bar{\nu}=0.13$ (B) and $\bar{\nu}=0.16$ (C) (see the 
stability    diagram in figure \ref{fig:diagstab}).   $e=0.6$ and $\phi = 
0.05$.}
  \label{fig:rd}
\end{figure}

\begin{figure}
  \centering   \psfrag{h}{\hspace{-5mm} $h$}
  \psfrag{lamdasurh}{\hspace{0mm} $\lambda_{c}/h$}
 % \psfrag{v}{\hspace{0mm} $\hat{v}$}
 % \psfrag{w}{\hspace{0mm} $\hat{w}$}
 % \psfrag{t}{\hspace{0mm} $\hat{T}$}
 % \psfrag{nu}{\hspace{0mm} $\hat{\nu}$}
 %\psfrag{nu}{\vspace{0mm} $\bar{\nu}$}
 %scale original 0.8
  \includegraphics[scale=0.6]{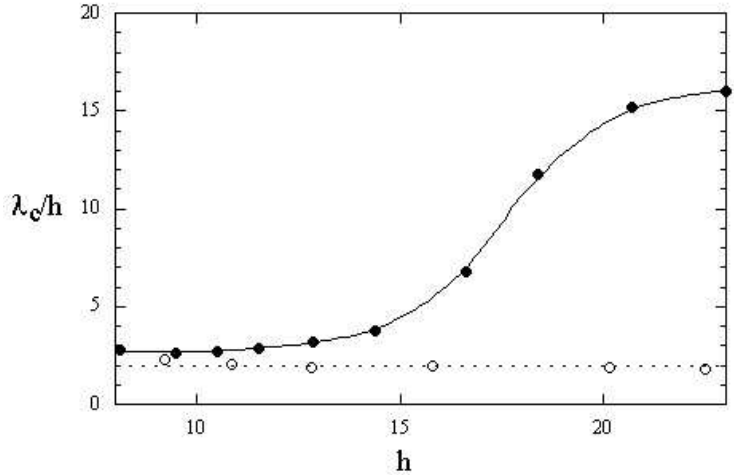}
  \caption{$\lambda_{c}/h$ ($\bullet$) as a function of $h$ , where $\lambda_{c}$   
is the wavelength selected by the instability along the marginal   
curve for the propagating instability. The white circles 
 ($\circ$) is the same curve obtained when the collisional dissipation
  $\gamma$ is artificially set to zero. ($e=0.6$, $\phi = 0.05$).}
    \label{fig:lambdaseuil}
  \end{figure}

The transition between the stationary instability and propagating 
modes appears clearly in figure \ref{fig:rd}. We have plotted 
in figure \ref{fig:rd} the growth rate $\Re \mathrm{e}(\sigma)$ and $\Im\mathrm{m}(\sigma)$ 
as a function of the wavenumber $k$, for the most dangerous mode at 
different locations in the stability diagram (A, B and C in figure 
\ref{fig:diagstab}). We keep the thickness constant ($h=12.8$) and 
increase the mean density from $\bar{\nu}=0.08$ to $\bar{\nu}=0.16$. 
One sees that the stationary mode in C appears from the collapse of 
the two conjugated propagating modes.  In figure \ref{fig:rd} we also 
observe that at the threshold of the instability, the first unstable 
mode occurs at a finite wavenumber $k_{c}$. We have systematically studied the critical 
wavelength $\lambda_{c} = 2\pi/k_{c}$ along the marginal curve. When the 
instability is stationary, the wavelength $\lambda_{c}$ is nearly 
constant and equal twice the thickness of the flow. By contrast, the 
wavelength  $\lambda_{c}$ strongly varies along the marginal curve 
for the propagating instability. In figure \ref{fig:lambdaseuil} we 
have plotted $\lambda_{c}/h$ as a function of $h$.  We observe that 
$\lambda_{c}$ is about $2.5$ the thickness $h$ of the flow as long as 
$h<15$ particle diameters but starts to increase for large thickness. 
For example, $h=21$ correspond to a wavelength 
 $\lambda_{c}=315$, which is $15$ times the thickness of the flow. We 
shall discuss later this behavior as the signature
of a change in the instability mechanism.   

\subsection{Eigenfunctions}

\begin{figure}
  \centering    \psfrag{z}{\hspace{0mm} $z$}
  \psfrag{u}{\hspace{0mm} $\hat{u}$}
  \psfrag{v}{\hspace{0mm} $\hat{v}$}
  \psfrag{w}{\hspace{0mm} $\hat{w}$}
  \psfrag{T}{\hspace{0mm} $\hat{T}$}
  \psfrag{nu}{\hspace{0mm} $\hat{\nu}$}
 %\psfrag{nu}{\vspace{0mm} $\bar{\nu}$}
 %scale original 0.8
  \includegraphics[scale=0.7]{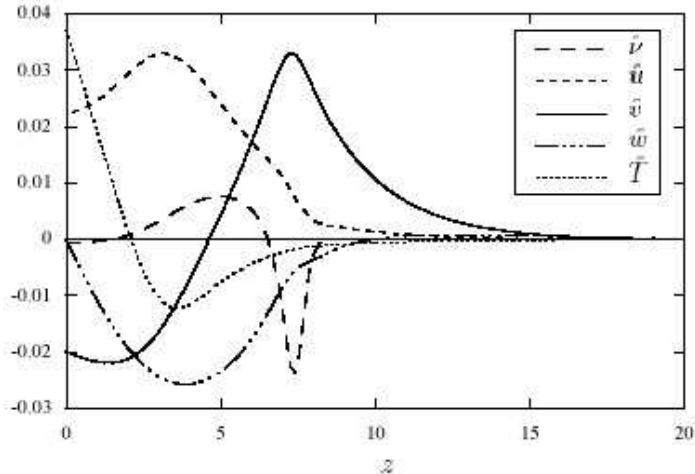}
  \caption{Real part of the eigenfunctions ($\hat{\nu}(z), 
\hat{u}(z), \hat{v}(z), \hat{w}(z), \hat{T}(z)$) corresponding to the 
stationary mode ($h=8$, $\bar{\nu}=0.23$, $k = 0.4$). ($e=0.6$, $\phi 
= 0.05$).}
    \label{fig:fp}
  \end{figure}

\begin{figure}
  \centering   \psfrag{z}{\hspace{-2mm} $z$}
  \psfrag{y}{\vspace{0mm} $y$}
  \psfrag{a}{\vspace{0mm} $(a)$}
  \psfrag{b}{\vspace{0mm} $(b)$}
  \psfrag{c}{\vspace{0mm} $(c)$}
  \psfrag{d}{\vspace{0mm} $(d)$}
  \psfrag{transverse velocity} {\vspace{0mm} transverse velocity}
  \psfrag{field} {\vspace{0mm} field}
  \psfrag{longitudinal velocity}{\vspace{0mm} longitudinal velocity}
  \psfrag{density}{\vspace{0mm} density}
  \psfrag{temperature}{\vspace{0mm} temperature}
  %scale original 0.7
  \includegraphics[scale=0.6]{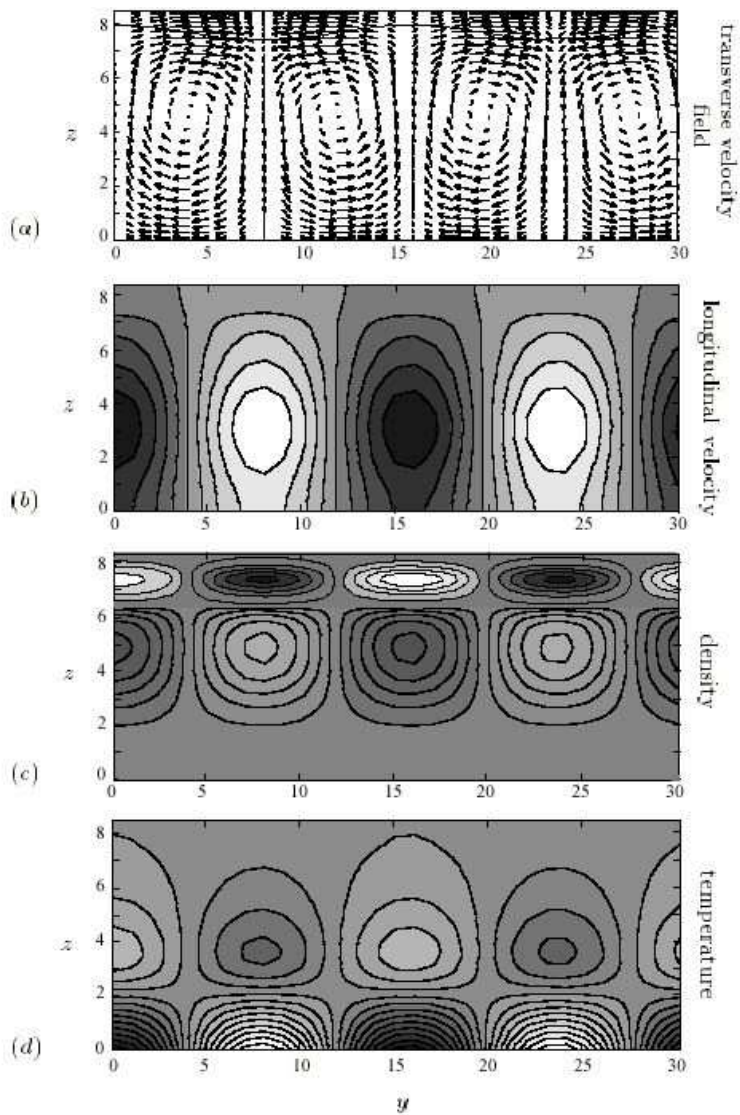}
  \caption{Perturbed fields corresponding to the    stationary mode 
($h=8$, $\bar{\nu}=0.23$, $k = 0.4$); (a) transverse velocity field, 
the bold solid line gives the free surface deformation; (b), (c), 
(d) contours of constant values for longitudinal velocity,  
density  and  temperature. The grey level is white for the most 
positive value and black for the most negative value. ($e=0.6$, $\phi 
= 0.05$).}
  \label{fig:vr}
  \end{figure}

The three-dimensional stability analysis of the basic flow gives also 
the
five eigenfunctions ($\hat{\nu}(z), \hat{u}(z), \hat{v}(z), 
\hat{w}(z), \hat{T}(z)$) for a given wavenumber $k$ and growth rate 
$\sigma $. In figure \ref{fig:fp}, we have plotted the real part of 
the eigenfunctions corresponding to the most amplified mode for 
($h=8$, $\bar{\nu}=0.23$). This mode is stationary. We observe that
the perturbation of density $\hat{\nu}(z)$ vanishes for $z>h$
whereas the  velocity and 
temperature perturbations  extend further and decay exponentially
 (see appendix B). From these eigenfunctions, we 
can simply recover the perturbed field $(\nu_{1}, u_{1}, v_{1}, 
w_{1}, T_{1})(y,z,t)$. For example, $\nu_{1}(y,z,t)$ is obtained
from the eigenfunction $\hat{\nu}(z)$ by the relation  $\nu_{1}(y,z,t) = 
\Re e [ \hat{\nu}(z)\exp (\sigma t + iky)]$ (see equation 4.2). In 
the following the perturbed field are plotted for $t=0$.
In figure \ref{fig:vr}, we present the perturbations of the flow 
corresponding to the most amplified stationary mode presented in 
figure \ref{fig:fp} 
($h=8$, $\bar{\nu}=0.23$). The pattern is typical of that 
observed in the stationary unstable region. Figure
\ref{fig:vr}(a)  presents the transverse  velocity field  
($v_{1}(y,z,0), w_{1}(y,z,0)$). We can see that the motion in the 
transverse plane consists in
counter-rotating vortices with a pair of vortices per wavelength 
$\lambda$.  This transverse flow is strongly correlated with the 
perturbation of the longitudinal velocity. Figure \ref{fig:vr}(b) 
shows the contours of constant value for the perturbed longitudinal 
velocity $u_{1}(y,z,0)$. The material going towards the plane is 
flowing faster in the slope direction than the material rising up to 
the free surface. Therefore, the 3D structure of the perturbed flow 
consists in 
longitudinal vortices with transverse variation of the longitudinal 
velocity. 

\begin{figure}
  \centering   \psfrag{y}{\hspace{0mm} $y$}
 % \psfrag{nu}{\vspace{0mm} $\bar{\nu}$}
 %scale original 0.9
  \includegraphics[scale=0.7]{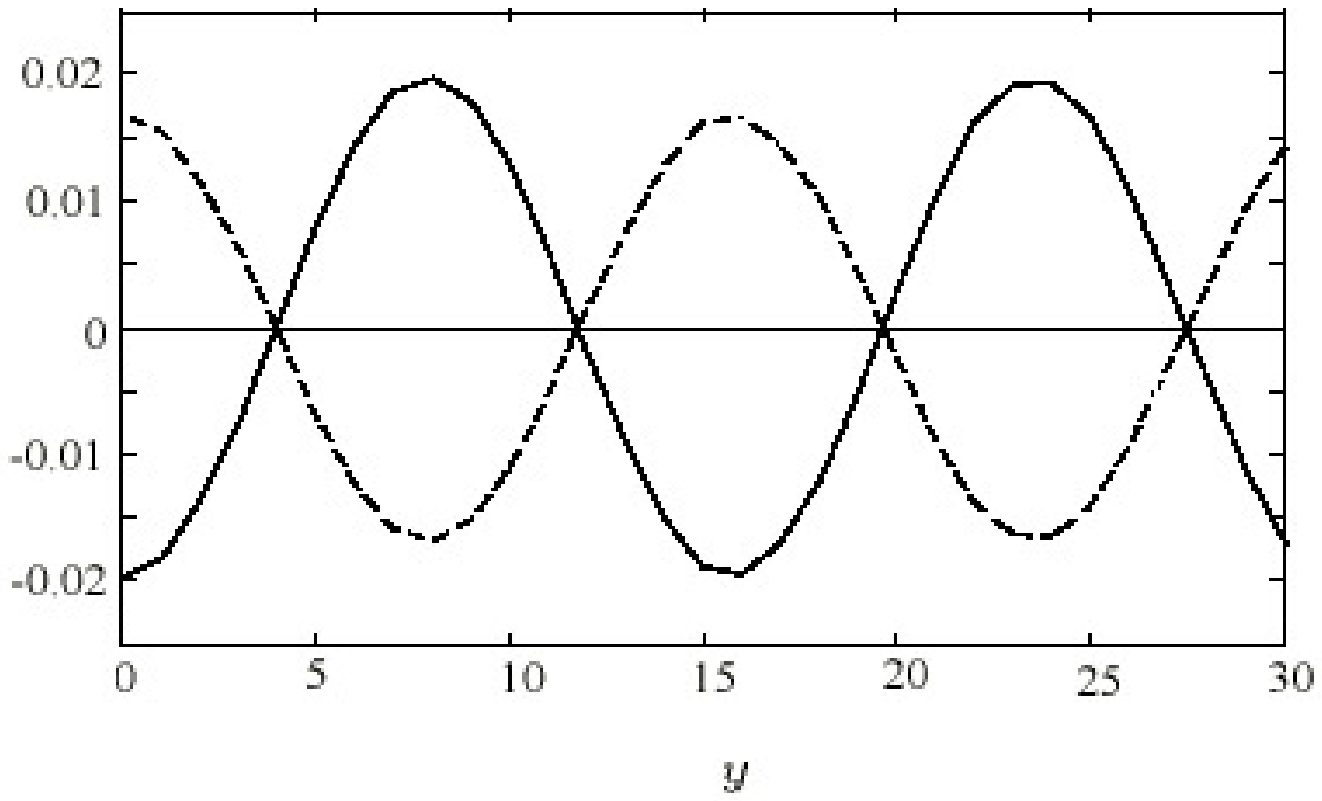}
  \caption{Perturbed depth averaged density $\mu(y) = 
\int_{0}^{z_{max}}\nu_{1}(y,z)\mathrm{d}z$ (solid line) as a function 
of  $y$ for the stationary unstable mode ($h=8$, $\bar{\nu}=0.23$, $k 
= 0.4$). The dotted line presents the transverse variations of the 
perturbed vertical velocity $w_{1}$ (arbitrary scale). The perturbed 
averaged density is higher where the flow is going downwards ($w_{1}<0$) and 
smaller when the flow is going upwards ($w_{1}>0$).  ($e=0.6$ and $\phi = 0.05$).}
  \label{fig:d}
  \end{figure}

The corresponding perturbations of density $\nu_{1}(y,z,0)$ and 
temperature $T_{1}(y,z,0)$ are given in figure \ref{fig:vr}(c) and 
(d). The perturbed density results from two effects: the advection of 
the basic flow by the transverse perturbed flow and the compressibility. 
In order to
compare more easily the density with the transverse flow, we have 
plotted in figure \ref{fig:d} the depth averaged density   $\mu(y) = 
\int_{0}^{z_{max}}\nu_{1}(y,z)\mathrm{d}z$  as a function of the 
transverse direction $y$ (solid line). The vertical 
velocity is given on the same plot (dotted line). We observe that
the average density is higher where the flow is going downwards and 
smaller when the flow is going upwards. The perturbation of 
density can also be interpreted in term of  free 
surface perturbation. If we define the free surface as being the surface where 
the total density $(\nu_{0}(z) + \nu_{1}(y,z,0))$ is constant and 
equal to $1\%$ of the maximum value of $\nu_{0}(z)$, we obtain the 
bold solid line in figure \ref{fig:vr}(a). One can see that the 
downward part
 of the flow corresponds to a through while the upward part of the flow 
corresponds to a crest.

\begin{figure}
  \centering   \psfrag{z}{\hspace{-2mm} $z$}
  \psfrag{y}{\vspace{0mm} $y$}
  \psfrag{a}{\vspace{0mm} $(a)$}
  \psfrag{b}{\vspace{0mm} $(b)$}
  \psfrag{c}{\vspace{0mm} $(c)$}
  \psfrag{d}{\vspace{0mm} $(d)$}
  \psfrag{transverse velocity}{\vspace{0mm} transverse velocity}
  \psfrag{field}{\vspace{0mm}field}
  \psfrag{longitudinal velocity}{\vspace{0mm} longitudinal velocity}
  \psfrag{density}{\vspace{0mm} density}
  \psfrag{temperature}{\vspace{0mm} temperature}
  \includegraphics[scale=0.6]{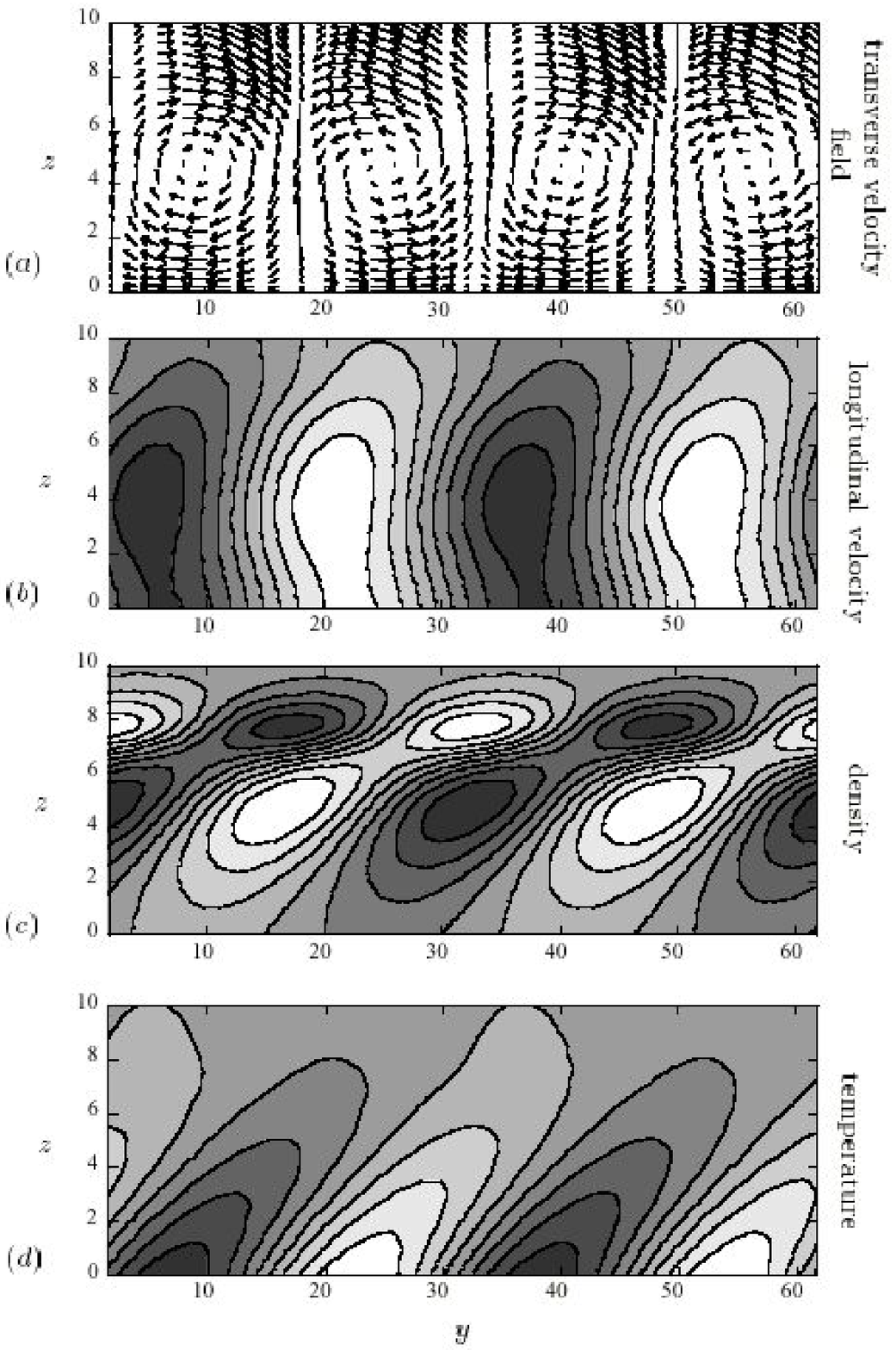}
  \caption{Perturbed fields corresponding to the    propagating  mode 
($h=10$, $\bar{\nu}=0.13$, $k = 0.2$); (a) transverse velocity 
field; (b), (c), (d) contours of constant values for  
longitudinal velocity,  density  and  temperature. ($e=0.6$, 
$\phi = 0.05$).}
  \label{fig:vi}
  \end{figure}

 Most of the features of the stationary modes are recovered for the 
propagating modes  except that the whole pattern is now drifting in 
the transverse direction, due to the non-zero phase velocity (see figure \ref{fig:vi}).
Qualitative differences exist since
the propagating modes lead to a phase shift between the 
eigenfunctions. The vortices are asymetric as shown in figure \ref{fig:vi}(a) 
and the transverse variations of density and longitudinal velocity are 
no longer in phase but slightly shifted.

\subsection{Influence of the parameters}

\begin{figure}
  \centering   \psfrag{z}{\hspace{0mm} $z$}
  \psfrag{y}{\vspace{0mm} $y$}
  %\psfrag{nu}{\vspace{1mm} $\nu_{0}(z)$}
  %\psfrag{T}{\vspace{1mm} $T_{0}(z)$}
  %scale original 0.6
  \includegraphics[scale=0.6]{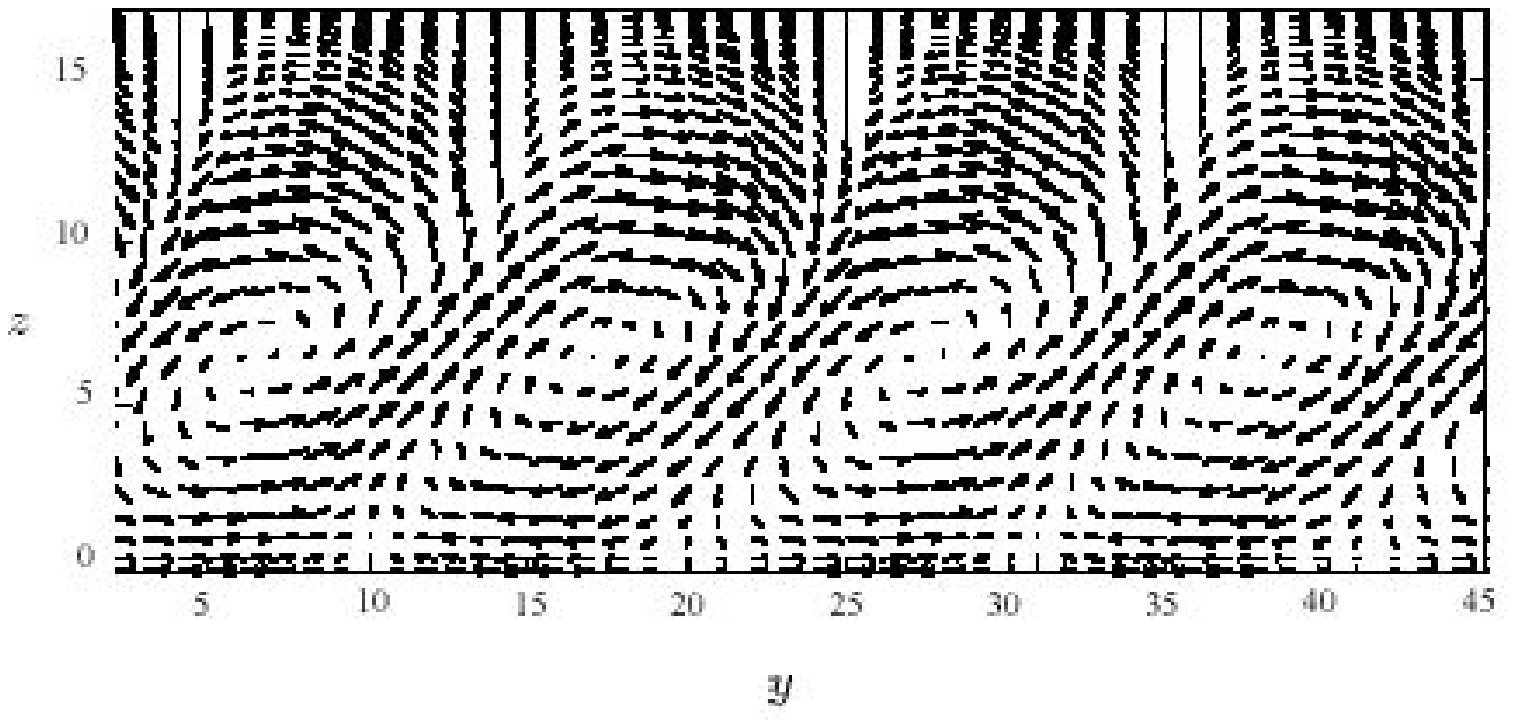}
  \caption{Typical transverse velocity field with $e=0.8$ and   
$\phi=0.12$ for the propagating instability ($h=17$,   
$\bar{\nu}=0.25$, $k=0.3$). }
  \label{fig:012}
  \end{figure}

The results of the stability analysis presented in the previous 
section have been performed using a given set of parameters $e=0.6$ 
and $\phi = 0.05$. Changing these parameters modifies the solutions 
for steady uniform flows and therefore modifies the results of the 
stability analysis. However, the main results of the previous section 
are not qualitatively changed in the range of parameters ($e$, 
$\phi$) where inverted density profiles are observed. For instance, 
we have perform the stability analysis using $e=0.8$ and $\phi = 
0.12$ and recover the instability and the formation of longitudinal 
vortices. The domain of propagating instability increases with higher 
values of $\phi$ and the vortices are more
tilted from the vertical  for the propagating unstable modes  (see figure 
\ref{fig:012}).

 However, an important difference with the case ($e=0.6$, 
$\phi=0.05$)  is that the  curve of marginal stability may overlap 
the domain of non-inverted density profiles. This means that for some 
values of $\phi$ and $e$, it is possible to find basic flows with 
non-inverted density profiles which are unstable. 
These non-inverted unstable flows are observed in a narrow range of 
the stability diagram for large thickness 
($h>30$) and small density ($\bar{\nu}<0.1$).  We will see in the 
next section that inelasticity is responsible of the instability of 
such non-inverted density profiles.

\section{Discussion}

\subsection{Instability mechanism and role of the inelasticity}
 In figure \ref{fig:diagstab}, the domain of non-inverted density 
profiles is shown together with the domain of instability. The 
correlation between the two regions suggests that inversion of the 
density profile plays an important role in the instability mechanism. 
We have checked the role of the gravity  by artificially modifying $\mathbf{g}$ in the 
linearized equations. In a wide range of thickness and mean density, 
increasing gravity increases the growth rate whereas decreasing 
gravity stabilizes the flow. In that case, the instability comes from 
the inversion of the density profile, which is due to the self-heating at the 
plane (``Rayleigh-B\'{e}nard'' type of instability mechanism). In fluid 
mechanics, the Rayleigh-B\'{e}nard instability is controlled at the 
threshold by a single
dimensionless parameter, the Rayleigh number  $R_{a} = g\rho \Delta 
\rho h^3/\eta K$. In a granular flow, it is difficult to define such 
a non-local control parameter since the Boussinesq approximation is 
far from being satisfied. Indeed, flow quantities (density, temperature) 
as well as transport coefficients (viscosities, conductivity) 
strongly vary inside the flow.

\begin{figure}
  \centering   \psfrag{h}{\hspace{0mm} $h$}
  \psfrag{nu}{\vspace{0mm} $\bar{\nu}$}
  %\psfrag{stable}{stable}
  %\psfrag{a}{A}
  %\psfrag{b}{B}
  %\psfrag{c}{C}   %\psfrag{density profiles}{density profiles}
  %\psfrag{stationary instability}{\vspace{0mm} stationary 
%instability}
  %\psfrag{oscillatory}{\vspace{0mm}   %propagating}
  %\psfrag{instability}{instability}
  %\psfrag{temperature}{\vspace{0mm} temperature}
  %scale original 0.6
  \includegraphics[scale=0.6]{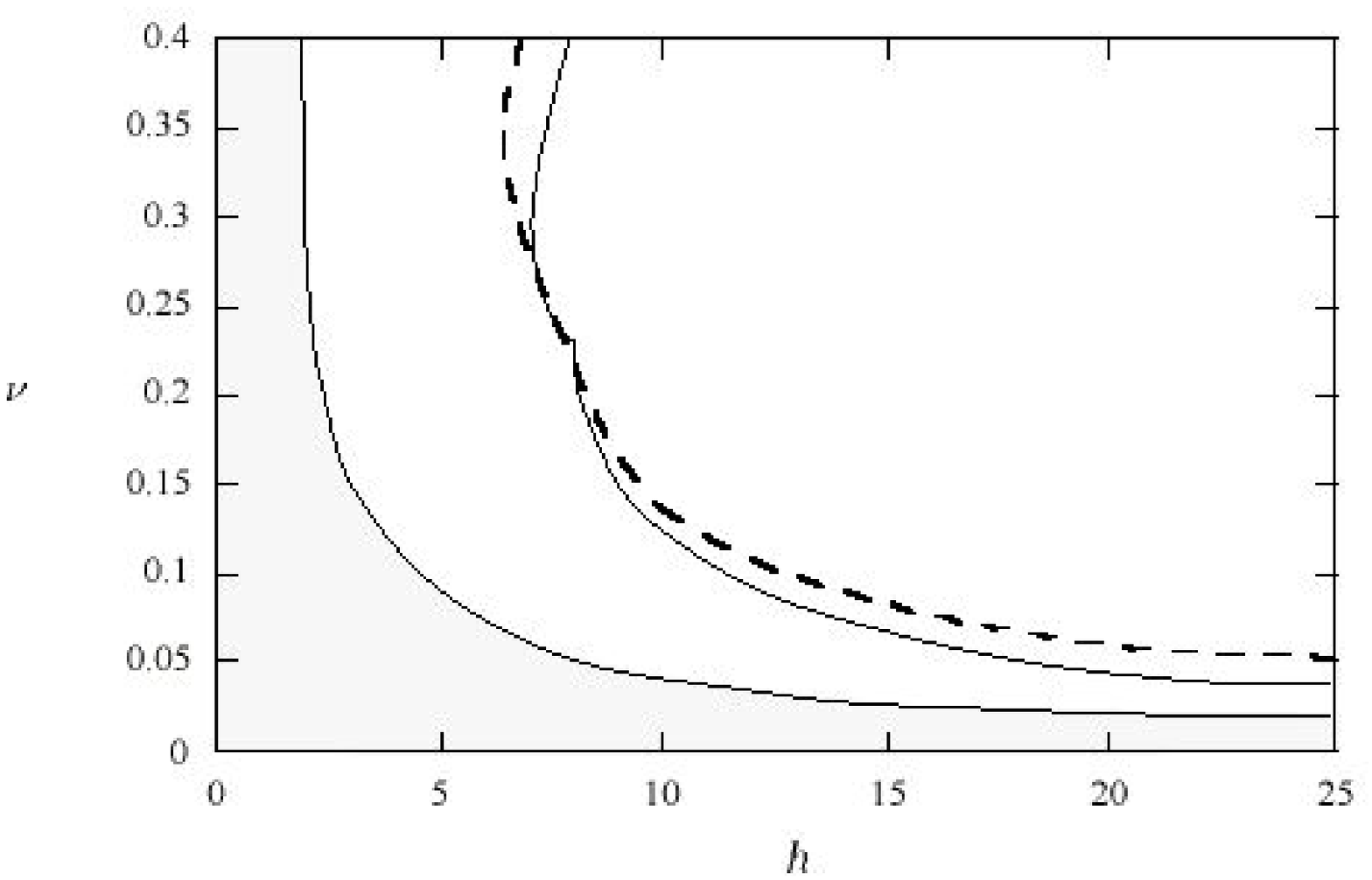}
  \caption{Role of inelasticity on the stability diagram. The solid 
line is the marginal curve with $\gamma \neq 0$ and the dotted line 
is the marginal curve with $\gamma = 0$ ($e=0.6$, $\phi = 0.05$).
   The grey zone is the domain of non-inverted density profiles.}
  \label{fig:ine}
  \end{figure}

 As pointed out in the previous 
section,  some flows with non-inverted density profiles are 
unstable (e.g. with $e=0.8$ and $\phi = 
0.12$). This means that gravity is not the only
destabilizing effect. Another well-known source of instability in 
granular flows is  
inelasticity. Studies on two-dimensional shear flows have shown that 
the dissipation due to inelastic collisions can lead to the formation 
of clusters (\cite{savage92}; \cite{wang97}; Alam \& Nott 1998; \cite{nott99}). In order to better understand 
the role of  dissipation in our problem,  we have 
performed the stability analysis by setting to zero the collisional 
rate of energy dissipation $\gamma$ in the linearized energy 
equation. In all the other terms, inelasticity is kept in order 
to study the same basic flow.  Figure \ref{fig:ine} 
presents on the same diagram the curve of marginal stability in the 
case $\gamma \neq 0$ and $\gamma = 0$ ($e=0.6$, $\phi=0.05$). We 
observe that the flow remains unstable in a wide range of parameters 
with $\gamma = 0$. This proves that inelasticity is not necessary to get 
an instability.
For thin and dense flows, inelasticity stabilizes the flow whereas for 
dilute flows inelasticity slightly lowers the threshold. The interesting point 
is that non-inverted density profiles are always stable without 
inelasticity, whatever the choice of parameters $e$ and $\phi$.
The dissipation also strongly influences  the wavelength selection  
 $\lambda_{c}$ at the threshold. In figure \ref{fig:lambdaseuil} we 
have plotted  $\lambda_{c}/h$
 as a function of $h$ both for  $\gamma \neq 0$ (black circles) and 
$\gamma = 0$  (white circles). When the dissipation is zero $\gamma = 
0$,
 the wavelength at the 
threshold scales with the thickness of the flow over the whole range 
of thickness ($\lambda_{c}\sim 2 h$). This implies that the  
 increase of $\lambda_{c}$ for large $h$ observed in the real 
 system ($\gamma \neq 0$) comes from 
inelasticity. 
 
This analysis suggests that both gravity and inelasticity contribute 
to the instability depending on the parameters. For thin and dense 
flows, gravity is the dominant destabilizing effect giving rise to a 
Rayleigh-B\'{e}nard kind of instability, whereas for large 
and dilute flows inelasticity is the principal ingredient of 
instability, giving rise to a clustering-like instability. However, it is not possible to separate the two effects 
since gravity and inelasticity are associated to the same unstable 
mode.

\subsection{Comparison with experimental observations}

The present study captures the principal features of the longitudinal 
vortices instability.
 First, our analysis has shown that steady uniform flows down 
inclined planes may be unstable under transverse perturbations. In the 
range of parameters of the experiment ($h=10-12$ grain diameters, $ 
\bar{\nu}=0.2-0.3$),  the physical origin of this instability is the 
inversion of the density profile, which is  induced by the  self-heating at the 
plane.  Secondly, the unstable flow consists in longitudinal 
vortices leading to surface deformation, transverse variations of 
longitudinal velocity and density in agreement with the experimental 
observations.
 As observed in the experiment, the longitudinal velocity is larger 
in the troughs (i.e. where the flow is going downwards) than in the crests 
(i.e. where the flow is going upwards). The same variation occurs with the 
density, troughs are dense and crests are dilute. Finally, the 
instability selects a wavelength which is always $2-3$ times the 
average thickness of the flow above the threshold, as observed in the 
experiment.

 It is more difficult to conclude about the relevance of 
this analysis in order to explain the phase velocity observed in the 
experiment. We have seen that depending on the control parameters 
($h$, $\bar{\nu}$), the growth rate of the instability may be complex 
i.e. $\sigma_{i}\neq 0$. This imaginary part of the growth rate is 
associated to a phase velocity $v_{\phi} = \sigma_{i}/k$, where $k$ is 
the wavenumber. A typical order of magnitude for $v_{\phi}$ is 
$(0.1-0.5) \sqrt{gd}$ which is a few percents of the chute velocity. 
This order of magnitude is compatible with the drift velocity 
observed in
the experiment. However, the theory predicts a transition between 
propagating and stationary instability, which is not observed in the experiment. Moreover, in the 
range of parameters of the experiment,  the theory predicts a stationary instability. 
The phase velocity observed experimentally is more likely
related to the very weak inclination of the pattern in the  
experiment. This inclination is not taken into account in this study since 
we have restricted our analysis to pure transverse perturbations 
($k_{x}=0$, $k_{y}\neq 0$). It would be interesting to investigate the 
stability of steady uniform flows when a small perturbation in the 
longitudinal direction ($k_{x}<<1$) is added to the $k_{y}$ 
perturbation. This should give rise to a phase velocity in the whole 
range of parameters.

The agreement between
 theory and  experiment is only qualitative.  The most 
important discrepancy between theory and experiment lies in the range 
of angles where the instability is predicted. With $e=0.6$ and 
$\phi=0.05$, the predicted angles are $\theta \sim 20^{\circ}$ while 
in the experiment the minimum angle in order to observe the 
instability is $\theta = 38^{\circ}$ with sand 0.25 mm diameter. The 
same order of magnitude is obtained in the experiment with monodisperse glass beads.  
In the theory, it is not possible to reach such inclination by 
varying the parameters ($e$, $\phi$) except for non-physical values 
of inelasticity ($e \sim 0$). Actually, this discrepancy between 
theory and experiment is not surprising since the standard 
kinetic theory is known to be limited. The kinetic theory has been 
first developed for quasi-elastic particles ($e\sim 1$) and rather dilute 
flows. Even under these conditions,  experimental studies of 
2D collisional granular 
flows down inclined planes have shown that the theory is not 
quantitative
(\cite{azanza99}). One of the main problem of the kinetic theory is 
the determination of the  boundary conditions at the rough plane. 
Experiments indicate that, even in the collisional regime, the 
material is structured near the plane (\cite{azanza99}). This local 
organization is not taken into account in the kinetic theory. Since 
boundary conditions control the production of granular temperature, 
it is not surprising that the theory fails in quantitatively 
describe steady uniform flows. Another difficulty of the kinetic 
theory lies in the treatment of strongly inelastic particules. For 
high inelasticity, additional terms should be introduced in the 
kinetic theory in order to take into account the lack of separation 
between the microscopic and macroscopic scales inherent to 
inelasticity (\cite{tan99} ; \cite{ sela98}). However, such 
additional terms complicate considerably the equations and are still 
the subject of active researches
(\cite{goldhirsch99}).   

\section{Conclusion}

In this paper, we have performed a three-dimensional stability analysis of 
rapid granular flows in the framework of the kinetic theory of granular 
gases.  We have shown that steady uniform flows down inclined planes can be unstable 
under transverse perturbations.  The structure of the unstable flow 
consists in longitudinal vortices with transverse variations of  free 
surface,  chute velocity and  density in  agreement with 
the experimental observations (\cite{forterre01}). Actually, the agreement is only 
qualitative. This is not surprising since the  experiments take place in a 
semi-dilute regime and use rather inelastic particles, which is beyond 
the domain of applicability of the simple kinetic theory used in the 
analysis. However, our study
 shows that the kinetic theory 
 is a relevant framework for the description of rapid granular flows. 
 The kinetic theory has revealed a new instability mechanism specific 
 to granular material: inelastic collisions trigger a self-induced 
 convection yielding longitudinal vortices in chute flows.

\begin{figure}
  \centering   \psfrag{x}{\hspace{1mm}$x$}
    \psfrag{y}{\hspace{1mm}$y$}
  %\psfrag{stable}{stable}
  %\psfrag{a}{A}
  %\psfrag{b}{B}
  %\psfrag{c}{C}   %\psfrag{density profiles}{density profiles}
  %\psfrag{stationary instability}{\vspace{0mm} stationary 
%instability}
  %\psfrag{oscillatory}{\vspace{0mm}   %propagating}
  %\psfrag{instability}{instability}
  %\psfrag{temperature}{\vspace{0mm} temperature}
  %scale original 0.6
  \includegraphics[scale=0.6]{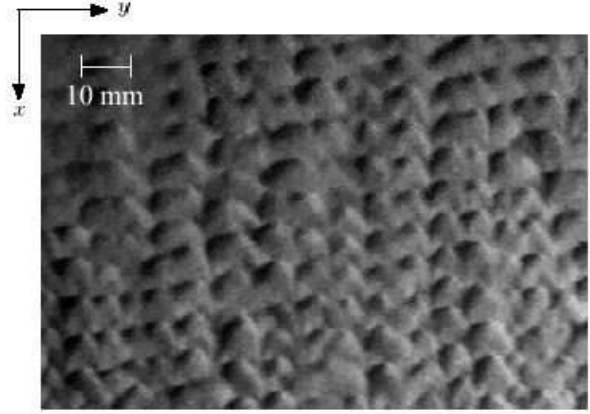}
  \caption{Top view of the free surface of the flow showing the   
formation of ``scales'' when the plane is strongly inclined (sand   
$0.25$ mm mean diameter, $\theta = 52^{\circ}$, $h_{g}=15$ mm). The
  picture is taken with a short shutter time of $1/10000 s$.}
  \label{fig:ecailles}
  \end{figure}

In classical fluid mechanics, Rayleigh-B\'{e}nard convection is the 
paradigm for pattern forming instabilities and the study of the 
transition to turbulence.  Therefore,
 an important question is 
whether the granular convection observed in our experiment  represents the 
starting point of a similar scenario towards more  complex structures.
Actually, we have observed in the experiment a new pattern when the plane is strongly 
inclined ($\theta >50^{\circ}$ with sand $0.25$ mm in mean diameter). 
Instead of longitudinal streaks, a regular square pattern looking as ``fish scales'' develops 
on the free surface, as shown in figure \ref{fig:ecailles}. Note that 
under such inclinations, a fast camera is necessary to capture this structure.
The appearance of this new pattern raises several issues. 
A first possibility is that for this
 range of parameters the square 
pattern represents the most unstable mode for the  primary 
instability, as for example observed in inclined layer convection 
(\cite{daniels00}). 
This could be investigated  by generalizing the present  
stability analysis to 
longitudinal modes i.e. $k_{x} \neq 0$ and $k_{y} \neq 0$.
A second possibility is that the ``scales'' result from a secondary 
instability of the longitudinal vortices, as a consequence of nonlinear 
effects. Such an evolution is well-documented  for classical fluid flows 
(Godr\`{e}che \& Manneville 1998) but still remains an open issue for granular flows.  
%In that case, the study of the scales would require a nonlinear analysis of the longitudinal 
%vortices.

  \acknowledgments
This work was supported by the French Ministry of 
Research and \'{E}ducation (ACI blanche $\# 2018$). We thank St\'{e}phane
 Le Diz\`{e}s and Jacques Liandrat for fruitful 
discussions.  
 \appendix

\section{}

In this Appendix we detail the main steps of the linear stability 
analysis presented in Section 4.1.

\subsection{linearization of the governing equations}
   The linearisation of the mass equation (\ref{eq:masscons}), the 
momentum equation (\ref{eq:pcons}) and the energy equation 
(\ref{eq:energycons}) around the basic flow ($\nu_{0}$, $u_{0}$, 
$T_{0}$) gives:
\begin{eqnarray}
			\frac{\partial \nu_{1}}{\partial t} & = & -\nu_{0}\left( 
\frac{\partial 			v_{1}}{\partial y} + \frac{\partial w_{1}}{\partial 
z}\right) - 			\frac{\mathrm{d}u_{0}}{\mathrm{d}z}w_{1}, 			
\label{eq1:nu}\\ 			\nu_{0}\left( \frac{\partial u_{1}}{\partial t} 
+ 			\frac{\mathrm{d}u_{0}}{\mathrm{d}z}w_{1}\right)
			& = & \nu_{1}\sin \theta + \frac{\partial 
{{\Sigma}_{xy}}_{1}}{\partial 			y} + \frac{\partial 
{{\Sigma}_{xz}}_{1}}{\partial z},
			\label{eq1:u}\\ 			\nu_{0} \frac{\partial v_{1}}{\partial t}& = & 
\frac{\partial 			{{\Sigma}_{yy}}_{1}}{\partial 			y} + 
\frac{\partial {{\Sigma}_{yz}}_{1}}{\partial z},
		    \label{eq1:v}\\
		    \nu_{0} \frac{\partial w_{1}}{\partial t}& = & 
-\nu_{1}\cos 		    \theta +\frac{\partial 			
{{\Sigma}_{zy}}_{1}}{\partial 			y} + \frac{\partial 
{{\Sigma}_{zz}}_{1}}{\partial z},
		    \label{eq1:w}\\
		    \frac{3}{2}\nu_{0}\left(\frac{\partial T_{1}}{\partial t} 
+ 		    \frac{\mathrm{d}T_{0}}{\mathrm{d}z}w_{1}\right)
		    & = & 
{{\Sigma}_{xz}}_{1}\frac{\mathrm{d}u_{0}}{\mathrm{d}z} 		    + 
{{\Sigma}_{xz}}_{0}\frac{\partial u_{1}}{\partial z}
		    + {{\Sigma}_{yy}}_{0}\frac{\partial v_{1}}{\partial y}
		    + {{\Sigma}_{zz}}_{0}\frac{\partial w_{1}}{\partial 		    
y}\nonumber \\
		    & & \hspace{37mm} - \frac{\partial {q_{y}}_{1}}{\partial y} - 
\frac{\partial 		    {q_{z}}_{1}}{\partial z} -\gamma_{1}.
			\label{eq1:T} \end{eqnarray}   where ${\mathbf{\Sigma}}_{1}$ is 
the stress tensor disturbance, ${\mathbf{q}}_{1}$ is the heat
 flux disturbance and $\gamma_{1}$ is the energy dissipation 
disturbance given by:
\begin{eqnarray*}
	{{\Sigma}_{xy}}_{1}& = &\eta_{0}\frac{\partial u_{1}}{\partial 	y}, 
\\
	{{\Sigma}_{xz}}_{1}& = &
	 \eta_{0}\frac{\partial u_{1}}{\partial z} + 	 
\eta_{1}\frac{\mathrm{d}u_{0}}{\mathrm{d}z}, \\
	 {{\Sigma}_{yy}}_{1} & = & -P_{1} + \left( 		    
\xi_{0}-\frac{2}{3}\eta_{0}\right)\left(\frac{\partial 
v_{1}}{\partial y}
		    +\frac{\partial w_{1}}{\partial z}\right) + 
2\eta_{0}\frac{\partial 		    v_{1}}{\partial y}, \\
	{{\Sigma}_{yz}}_{1} & = & \eta_{0}\left(\frac{\partial 
v_{1}}{\partial z}
		    +\frac{\partial w_{1}}{\partial y}\right), \\
		    {{\Sigma}_{zz}}_{1} & = & -P_{1} + \left( 		    
\xi_{0}-\frac{2}{3}\eta_{0}\right)\left(\frac{\partial 
v_{1}}{\partial y}
		    +\frac{\partial w_{1}}{\partial z}\right) + 
2\eta_{0}\frac{\partial 		    w_{1}}{\partial z},\\
		    {q_{y}}_{1} & = &K_{0} \frac{\partial 		    T_{1}}{\partial y}, 
\\
		    {q_{z}}_{1} & = & K_{0}\frac{\partial T_{1}}{\partial z} + 	
K_{1}\frac{\mathrm{d}T_{0}}{\mathrm{d}z}.   \end{eqnarray*}
Here the pressure disturbance $P_{1}$, the viscosity disturbance 
$\eta_{1}$, the conductivity disturbance $K_{1}$ and the energy 
dissipated disturbance $\gamma_{1}$ are linearized functions of 
$\nu_{1}$ and $T_{1}$ which are given in table 2.\\
The system of equations (\ref{eq1:nu})-(\ref{eq1:T}) must be 
considered together with the boundary conditions for the disturbances 
$(\nu_{1}(z), {u}_{1}(z),  {v}_{1}(z), {w}_{1}(z), T_{1}(z))$. At the 
plane, the boundary  conditions   are written  as:
\begin{eqnarray}
			w_{1} & = & 0\hspace{5mm} \mbox{at $z=0$}, 			
\label{bc1plan:w}\\ 			\eta_{0} \frac{\partial u_{1}}{\partial z} + 
\eta_{1}
			\frac{\mathrm{d}u_{0}}{\mathrm{d}z}
			& = &  {\eta_{0}}^{\ast}u_{1} + {\eta_{1}}^{\ast}u_{0}\hspace{5mm} 
\mbox{at $z=0$},
			\label{bc1plan:u}\\ 			\eta_{0} \left( \frac{\partial 
v_{1}}{\partial z} + \frac{\partial 			w_{1}}{\partial y} \right) & = 
& {\eta_{0}}^{\ast}v_{1}\hspace{5mm} \mbox{at $z=0$},
		    \label{bc1plan:v}\\
		    -K_{0}\frac{\partial T_{1}}{\partial z} - 		    
K_{1}\frac{\mathrm{d}T_{0}}{\mathrm{d}z}
		    & = & \eta_{1}
			\frac{\mathrm{d}u_{0}}{\mathrm{d}z}u_{0} + 
\eta_{0}u_{0}\frac{\partial u_{1}}{\partial z}
			+ \eta_{0}\frac{\mathrm{d}u_{0}}{\mathrm{d}z}u_{1} + 			
{K_{1}}^{\ast}\hspace{5mm} \mbox{at $z=0$,}
			\label{bc1plan:T}
			 \end{eqnarray}
where ${\eta_{1}}^{\ast}$ and ${K_{1}}^{\ast}$ are given in table 2. 
The first condition (\ref{bc1plan:w}) simply expresses that the plane 
is rigid. The last three conditions 
(\ref{bc1plan:u})-(\ref{bc1plan:T}) are the boundary conditions 
(\ref{bc:stress}) and (\ref{bc:energy}), which are linearized around 
the basic flow ($\nu_{0}$, $u_{0}$, $T_{0}$). At infinity, we assume  
the disturbances to vanish:
  \begin{equation}
  	\nu_{1}, u_{1}, v_{1}, w_{1}, T_{1}\rightarrow 0;   	
\hspace{10mm}\mbox{when}\;z\rightarrow \infty.
  	\label{bc1:infini}
  \end{equation} 
  
    \begin{table}
\centering
\begin{eqnarray*}
    P_{1}   & = &     {f_{1}}'({\nu}_{0}) T_{0}\nu_{1} 
+f_{1}({\nu}_{0})T_{1} =     a_{0}\nu_{1} + b_{0}T_{1}\\
     \eta_{1} & = &     {f_{2}}'({\nu}_{0}) 
{T_{0}}^{\frac{1}{2}}\nu_{1} + \frac{1}{2}f_{2}({\nu}_{0}) 
{T_{0}}^{-\frac{1}{2}}T_{1}
    = c_{0}\nu_{1} + d_{0}T_{1}\\
    K_{1}  & = &     {f_{3}}'({\nu}_{0}) {T_{0}}^{\frac{1}{2}}\nu_{1} 
+ \frac{1}{2}f_{3}({\nu}_{0}) {T_{0}}^{-\frac{1}{2}}T_{1}
    = e_{0}\nu_{1} + h_{0}T_{1}\\
    \gamma_{1} & = &     {f_{5}}'({\nu}_{0}) 
{T_{0}}^{\frac{3}{2}}\nu_{1} + \frac{3}{2}f_{5}({\nu}_{0}) 
{T_{0}}^{\frac{1}{2}}T_{1}
    = l_{0}\nu_{1} + m_{0}T_{1}\\
    {\eta_{1}}^{\ast} & = &     \phi {f_{6}}'({\nu}_{0}) 
{T_{0}}^{\frac{1}{2}}\nu_{1} + \frac{1}{2}\phi f_{6}({\nu}_{0}) 
{T_{0}}^{-\frac{1}{2}}T_{1}
    = n_{0}\nu_{1} + p_{0}T_{1}\\
     {\gamma_{1}}^{\ast}  & = &     (1-e^2) {f_{7}}'({\nu}_{0}) 
{T_{0}}^{\frac{3}{2}}\nu_{1} + \frac{3}{2}(1-e^2) 
f_{7}({\nu}_{0})     {T_{0}}^{\frac{1}{2}}T_{1}= q_{0}\nu_{1} + 
r_{0}T_{1}
    \end{eqnarray*}
\caption{Pressure disturbance $P_{1}$,  viscosity disturbance 
$\eta_{1}$, conductivity disturbance $K_{1}$ and  energy dissipated 
disturbance $\gamma_{1}$}
	\label{transportdisturbance}
	\end{table}

\subsection{Matrices of the eigenvalue problem}

The non-zero elements of the coefficient matrices in equations 
(\ref{eq1:final}) and (\ref{bc1plan:final}) are given below. The 
definition of the functions of the basic flow 
($a_{0}$,\ldots,$r_{0}$) are given in table 2.

 \begin{eqnarray*}
L_{22} & = & \eta_{0},\\
L_{33} &=& \eta_{0},\\
L_{44}& =& \xi_{0} +\frac{4}{3}\eta_{0},\\
L_{55} &=& K_{0},\\
M_{14}& =& \nu_{0},\\
M_{21} &=& c_{0}\frac{\mathrm{d}u_{0}}{\mathrm{d}z},\\
M_{22} &=& \frac{\mathrm{d}\eta_{0}}{\mathrm{d}z},\\
M_{25}& =& d_{0}\frac{\mathrm{d}u_{0}}{\mathrm{d}z},\\
M_{33}& =& \frac{\mathrm{d}\eta_{0}}{\mathrm{d}z},\\
M_{34} &=& ik\left( \xi_{0} +\frac{1}{3}\eta_{0}\right),\\
M_{41}& =& -a_{0},\\
M_{43} &=& ik\left( \xi_{0} +\frac{1}{3}\eta_{0}\right),\\
M_{44} &=& \frac{\mathrm{d}\xi_{0}}{\mathrm{d}z} 
+\frac{4}{3}\frac{\mathrm{d}\eta_{0}}{\mathrm{d}z},\\
M_{45}& =& -b_{0},\\
M_{51} &=& e_{0}\frac{\mathrm{d}T_{0}}{\mathrm{d}z},\\
M_{52} &=& 2\eta_{0}\frac{\mathrm{d}u_{0}}{\mathrm{d}z},\\
M_{54}& =& -P_{0},\\
M_{55}& =& \frac{\mathrm{d}K_{0}}{\mathrm{d}z} + 
h_{0}\frac{\mathrm{d}T_{0}}{\mathrm{d}z},\\
N_{11} &=& \sigma,\\
N_{13}& =& ik\nu_{0},\\
N_{14}& =& \frac{\mathrm{d}u_{0}}{\mathrm{d}z},\\
N_{21} &=& \sin \theta + c_{0}\frac{\mathrm{d^2}u_{0}}{\mathrm{d}z^2} 
+
\frac{\mathrm{d} 
c_{0}}{\mathrm{d}z}\frac{\mathrm{d}u_{0}}{\mathrm{d}z},\\
N_{22}& =& -\sigma \nu_{0} -k^2 \eta_{0},\\
N_{24}& =& -\nu_{0}\frac{\mathrm{d}u_{0}}{\mathrm{d}z},\\
N_{25} &=& d_{0}\frac{\mathrm{d^2}u_{0}}{\mathrm{d}z^2} +
\frac{\mathrm{d} 
d_{0}}{\mathrm{d}z}\frac{\mathrm{d}u_{0}}{\mathrm{d}z},\\
N_{31} &=& -ika_{0},\\
N_{33} &=& -\sigma \nu_{0} -k^2\left( \xi_{0} 
+\frac{4}{3}\eta_{0}\right),\\
N_{34}& =& ik\frac{\mathrm{d}\eta_{0}}{\mathrm{d}z},\\
N_{35} &=& -ikb_{0},\\
N_{41}& = &-\cos \theta - \frac{\mathrm{d}a_{0}}{\mathrm{d}z},\\
N_{43} &=& ik\left( \frac{\mathrm{d}\xi_{0}}{\mathrm{d}z} 
-\frac{2}{3}\frac{\mathrm{d}\eta_{0}}{\mathrm{d}z}\right),\\
N_{44}& =& -\sigma \nu_{0}-k^2 \eta_{0},\\
N_{45} &=& -\frac{\mathrm{d}b_{0}}{\mathrm{d}z},\\
N_{51} &=& c_{0}{\left(\frac{\mathrm{d}u_{0}}{\mathrm{d}z}\right)}^2 
+ e_{0}\frac{\mathrm{d^2}T_{0}}{\mathrm{d}z^2} +
\frac{\mathrm{d} 
e_{0}}{\mathrm{d}z}\frac{\mathrm{d}T_{0}}{\mathrm{d}z} - l_{0},\\
N_{53}& = &-ikP_{0},\\
N_{54}& =& -\frac{3}{2}\nu_{0}\frac{\mathrm{d}T_{0}}{\mathrm{d}z},\\
N_{55}& =& -\frac{3}{2}\nu_{0}\sigma -k^2 K_{0}+ 
d_{0}{\left(\frac{\mathrm{d}u_{0}}{\mathrm{d}z}\right)}^2 + 
h_{0}\frac{\mathrm{d^2}T_{0}}{\mathrm{d}z^2} +
\frac{\mathrm{d} 
h_{0}}{\mathrm{d}z}\frac{\mathrm{d}T_{0}}{\mathrm{d}z} - m_{0},\\
Q_{22} & = & \eta_{0},\\
Q_{33} & = & \eta_{0},\\
Q_{52} & = & -\eta_{0}u_{0},\\
Q_{55} & = & -K_{0},\\
R_{14} & = & 1,\\
R_{21} & = & c_{0}\frac{\mathrm{d}u_{0}}{\mathrm{d}z} - u_{0}n_{0},\\
R_{22} & = & -{\eta_{0}}^{\ast},\\
R_{33} & = & -{\eta_{0}}^{\ast},\\
R_{34} & = & ik\eta_{0},\\
R_{51} & = & -e_{0}\frac{\mathrm{d}T_{0}}{\mathrm{d}z} - 
c_{0}u_{0}\frac{\mathrm{d}u_{0}}{\mathrm{d}z} + q_{0},\\
R_{52} & = & -\eta_{0}\frac{\mathrm{d}u_{0}}{\mathrm{d}z},\\
R_{55} & = & -h_{0}\frac{\mathrm{d}T_{0}}{\mathrm{d}z} - 
d_{0}u_{0}\frac{\mathrm{d}u_{0}}{\mathrm{d}z} + r_{0}.
\end{eqnarray*}
\section{}

Here we give the asymptotic behaviour of the disturbances  
($\hat{\nu}$, $\hat{u}$, $\hat{v}$, $\hat{w}$, $\hat{T}$) when $z$ is 
much larger than the characteristic thickness of the basic flow. The 
linearized equations (\ref{eq1:final}) which govern the disturbances 
can be written as
\begin{eqnarray}
			\sigma \hat{\nu} & = & -\nu_{0}\left( ik\hat{v} + 			
\frac{\mathrm{d}\hat{w}}{\mathrm{d}z}\right) - 
\frac{\mathrm{d}u_{0}}{\mathrm{d}z}\hat{w}, 			
\label{eq1modal:nu}\\ 
\nu_{0}\left(\sigma \hat{u} + 
\frac{\mathrm{d}u_{0}}{\mathrm{d}z}\hat{w}\right)
			& = & \hat{\nu}\sin \theta -k^2{\eta}_{0}\hat{u} + 			
\frac{\mathrm{d}}{\mathrm{d}z}\left(\eta_{0}\frac{\mathrm{d}\hat{u}}{\mathrm{d}z}
		+ \hat{\eta}\frac{\mathrm{d}u_{0}}{\mathrm{d}z}\right),
			\label{eq1modal:u}\\ 
\nu_{0}\sigma \hat{v} & = & -ik\hat{P} +
			(\xi_{0}-\frac{2}{3}\eta_{0})\left(-k^2\hat{v} + ik 			
\frac{\mathrm{d}\hat{w}}{\mathrm{d}z}\right) - 			
2\eta_{0}k^2\hat{v}\nonumber \\
			& & + \frac{\mathrm{d}}{\mathrm{d}z}\left( 			
\eta_{0}\frac{\mathrm{d}\hat{v}}{\mathrm{d}z} + ik\eta_{0} 			
\hat{w}\right),
		    \label{eq1modal:v}\\
\nu_{0}\sigma \hat{w} & = & -\hat{\nu}\cos \theta - 		    
\frac{\mathrm{d}\hat{P}}{\mathrm{d}z} +\eta_{0}\left(ik 
\frac{\mathrm{d}\hat{v}}{\mathrm{d}z}
		    -k^2\hat{w}\right) + 		    \frac{\mathrm{d}}{\mathrm{d}z}\left[ 
\left( 		    \xi_{0}-\frac{2}{3}\eta_{0}\right)
		    \left( ik\hat{v} + \frac{\mathrm{d}\hat{w}}{\mathrm{d}z} 		    
\right) \right]\nonumber \\
		    & &  + 2 		    \frac{\mathrm{d}}{\mathrm{d}z}\left( 
\eta_{0}\frac{\mathrm{d}\hat{w}}
		    {\mathrm{d}z}\right),
		    \label{eq1modal:w}\\
\frac{3}{2}\nu_{0}\left(\sigma \hat{T} + 
\frac{\mathrm{d}T_{0}}{\mathrm{d}z}\hat{w}\right)
		    & = & \left( \eta_{0}\frac{\mathrm{d}\hat{u}}{\mathrm{d}z}+ 
\hat{\eta}\frac{\mathrm{d}u_{0}}{\mathrm{d}z}\right)
		    \frac{\mathrm{d}u_{0}}{\mathrm{d}z} + 
\eta_{0}\frac{\mathrm{d}u_{0}}{\mathrm{d}z}\frac{\mathrm{d}\hat{u}}{\mathrm{d}z}
		    -P_{0}\left( ik\hat{v} + 			
\frac{\mathrm{d}\hat{w}}{\mathrm{d}z}\right) -k^2 
K_{0}\hat{T}\nonumber \\
			& & +
\frac{\mathrm{d}}{\mathrm{d}z}\left(K_{0}\frac{\mathrm{d}\hat{T}}{\mathrm{d}z}
+ \hat{K}\frac{\mathrm{d}T_{0}}{\mathrm{d}z}\right) -\hat{\gamma}.
\label{eq1modal:T} 
\end{eqnarray}
In order to obtain the 
asymptotic expression of these equations, one has to know the 
asymptotic behaviour of the basic flow. Equation (\ref{eq0:nu0}) 
shows that the zero-order density $\nu_{0}$ decays exponentially to 
zero as 
$$
\nu_{0} \rightarrow \exp\left(-\frac{\cos \theta}{T_{0}}z\right).	
$$
when $z$ is large. Using equations (\ref{eq0:nu0})-(\ref{eq0:T0}), 
the asymptotic behaviour of the basic flow is given by 
\begin{eqnarray}
\frac{\mathrm{d}u_{0}}{\mathrm{d}z}\sim \nu_{0};& & 
\frac{\mathrm{d^2}u_{0}}{\mathrm{d}z^2}\sim \nu_{0},
\label{as01}\\
\frac{\mathrm{d}T_{0}}{\mathrm{d}z}\rightarrow 0;& & 
\frac{\mathrm{d^2}T_{0}}{\mathrm{d}z^2}\rightarrow 0,
\label{as02}\\
\eta_{0}\rightarrow C;& & 
\frac{\mathrm{d}\eta_{0}}{\mathrm{d}z}\rightarrow 0,
\label{as03}\\
K_{0}\rightarrow D;& & \frac{\mathrm{d}K_{0}}{\mathrm{d}z}\rightarrow 
0,
\label{as04}\\
P_{0}\sim \nu_{0},& &
\label{as05}\\
\xi_{0}\sim {\nu_{0}}^2;& & \frac{\mathrm{d}\xi_{0}}{\mathrm{d}z}\sim 
{\nu_{0}}^2,\label{as06}
\end{eqnarray}
Considering (\ref{as01})-(\ref{as06}) into the linearized equations 
(\ref{eq1modal:nu})-(\ref{eq1modal:T}) gives an asymptotic behaviours 
for $z$ which is much larger than  the characteristic thickness of 
the basic flow.  From  equation (\ref{eq1modal:nu}), one  shows that 
the perturbed density decays on the same length scale as the basic 
flow:  \begin{equation}
\hat{\nu}\sim \nu_{0}.
\end{equation} Therefore, the asymptotic limit of equations 
(\ref{eq1modal:nu})-(\ref{eq1modal:T})
is given by
\begin{eqnarray}
\frac{\mathrm{d^2}\hat{u}}{\mathrm{d}z^2}-k^2\hat{u}& = &
0,
\label{as1}\\
\frac{\mathrm{d^2}\hat{v}}{\mathrm{d}z^2}-\frac{4}{3}k^2\hat{v}
+\frac{1}{3} ik\frac{\mathrm{d}\hat{w}}{\mathrm{d}z}& = &
0,
\label{as2}\\
\frac{1}{3} ik\frac{\mathrm{d}\hat{v}}{\mathrm{d}z} + 
\frac{4}{3}\frac{\mathrm{d^2}\hat{w}}{\mathrm{d}z^2}-k^2\hat{w}
& = &
0,
\label{as3}\\
\frac{\mathrm{d}\hat{T}}{\mathrm{d}z}-k^2\hat{T}& = &
0.
\label{as4}
\end{eqnarray}
These equations  are easily solved and give
$$
\begin{array}{cc}
\left(\begin{array}{c}
\hat{u}\\
\hat{v}\\
\hat{w}\\
\hat{T}
\end{array}\right) & \sim \exp (-kz).
\end{array}
$$
when $z$ is much larger than the characteristic thickness of the basic 
flow.
%\newpage

         \newpage

    \end{document}